\documentclass[11pt]{article}
\pdfoutput=1 

\usepackage{jheppub} 

\usepackage[T1]{fontenc} 
\usepackage{graphicx}
\usepackage[numbers]{natbib}
\usepackage{url,amsmath,amssymb,amsfonts}
\usepackage{subcaption}
\usepackage{multirow}
\usepackage{placeins}
\usepackage{makecell}
\usepackage{xcolor}
\newcommand{\nc}{\newcommand}
\nc{\gev}{\;\mathrm{GeV}}
\newcommand{\br}{{\mathrm{Br}}}
\newcommand{\inv}{{\mathrm{invisible}}}

\title{Exotic Higgs Decays}

\author[a]{Mar\'{i}a Cepeda}
\author[b]{Stefania Gori}
\author[c]{Verena Mart\'{i}nez Outschoorn}
\author[d]{Jessie Shelton}

\affiliation[a]{Department of Basic Research, CIEMAT,\\Madrid, Spain, 28040}
\affiliation[b]{Santa Cruz Institute for Particle Physics and Department of Physics, University of California, Santa Cruz,\\1156 High Street, Santa Cruz, CA 95064}
\affiliation[c]{Department of Physics, University of Massachusetts at Amherst\\Amherst, MA, USA, 01003}
\affiliation[d]{Illinois Center for the Advanced Studies of the Universe, Department of Physics, University of Illinois at Urbana-Champaign,\\Urbana, IL, USA, 61801}

\emailAdd{cepeda@cern.ch}
\emailAdd{sgori@ucsc.edu}
\emailAdd{vimartin@umass.edu}
\emailAdd{sheltonj@illinois.edu}

\abstract{Exotic decays of the Standard Model-like Higgs boson into beyond-the-Standard Model particles are predicted in a wide range of well-motivated theories.  The enormous samples of Higgs bosons that have been and will be produced at the Large Hadron Collider thus constitute one of the key discovery opportunities at that facility, particularly in the upcoming high-statistics high-luminosity run.  Here we review recent theoretical work on models that predict or accommodate exotic Higgs decays, the status of current experimental searches, and look forward to future capabilities at dedicated Higgs factories and beyond.}

\begin{document} 
\maketitle
\flushbottom

\section{Introduction}
\label{sec:intro}

With the discovery of a Standard Model (SM)-like Higgs boson \cite{ATLAS:2012yve,CMS:2013btf}, experimental probes of the electroweak (EW) scale have entered a new era.  The stunning validation of SM predictions at the Large Hadron Collider (LHC) and past collider experiments has upended decades of thought about the hierarchy problem, the origins of dark matter (DM), and the nature of new physics, and puts precision probes of the SM Higgs at the heart of the experimental discovery program for the foreseeable future.

One of the major outstanding discovery opportunities in this program is the search for exotic decays of the Higgs boson, where the Higgs decays to light new particles.  Rare decays of SM particles are frequently an excellent place to search for the footprints of new physics with suppressed couplings to the SM, which has motivated searches for e.g. rare decays of the $Z$ boson, of $B$, $K$, and $\pi$ mesons, and so on.  Among the SM particles, however, the Higgs stands out as a special case given the ease with which it can mediate interactions with SM-singlet new particles. The Higgs portal operator $|H|^2$, along with the hypercharge field strength $B_{\mu\nu}$, is the lowest-dimension total singlet operator that can be constructed out of SM fields, and thus will frequently mediate the leading interaction between SM and beyond the Standard Model (BSM) physics at the energy scales testable at the LHC. Thus exotic Higgs decays can easily be the leading signal of a wide variety of theories. Moreover, we have many theoretical motivations to suspect that new physics might couple preferentially to the Higgs, as the Higgs sits at the heart of many outstanding mysteries in particle physics. 

Additionally, the Higgs has an unusually tiny width since all of its SM decay modes are accidentally small, suppressed by  small Yukawa couplings, multibody phase space, or loop factors. For this reason,  
even tiny couplings to new physics can lead to experimentally interesting branching ratios into exotic final states.    With the new discovery landscape provided by the copious Higgs samples newly available at the LHC and at future Higgs factories including the high-luminosity (HL) LHC, a broad and generic search for exotic Higgs decays represents one of the leading discovery opportunities for new physics in the near future.
 
Exotic Higgs decays have long been appreciated as potential signals of new physics, beginning with  the pioneering work of Suzuki and Shrock \cite{Shrock:1982kd}.   A large body of work on exotic Higgs decays has been motivated by solutions to the hierarchy problem, particularly in the context of extended SUSY models 
(though see also composite Higgs models) 
as well as Higgs portal dark matter (see the 
paper \cite{Curtin:2013fra} for details about both models and signatures).   
In this review article, we focus on theoretical motivations for exotic Higgs decays that have been substantially developed since \cite{Curtin:2013fra} and the major advances in the experimental search for these signatures.

After reviewing theoretical motivations  
in section~\ref{sec:theory}, we detail three simple benchmark scenarios in section~\ref{sec:benchmarks} that realize features of many of the best-developed theories predicting exotic decays.   Current experimental searches are summarized in section~\ref{sec:exp}, including interpretations of both prompt and long-lived searches in our benchmark scenarios.  In section~\ref{sec:future} we survey future prospects, and we conclude in section~\ref{sec:conclusions}.

\section{Theoretical motivations}
\label{sec:theory}

There are many reasons to expect new physics to lie near the electroweak scale.  New  weak-scale states might be directly tied to the electroweak symmetry-breaking (EWSB) process, and/or responsible for controlling the stability of the electroweak scale relative to the Planck scale.  Thermal relic DM independently motivates new particles with mass at or below the weak scale.  Many models inspired by these three directions directly involve the Higgs boson and often naturally predict Higgs decays as one of the leading terrestrial signatures of the model, as this section will discuss in detail.  However, while these three topics most straightforwardly relate Higgs decays to long-standing mysteries in particle physics, they by no means exhaust the scenarios that can motivate exotic Higgs decay modes. More broadly, the SM Higgs provides a powerful window onto the possible existence of low-mass SM singlets.  Motivations for new light particles 
can also include the strong CP problem, which strongly motivates the existence of axions or axion-like-particles (ALPs), and the observed baryon-anti-baryon asymmetry, which cannot be generated in the SM.   Outstanding experimental anomalies, such as the muon $g-2$, may also hint at light BSM physics. 
Exotic decays of the SM Higgs boson can provide a unique and powerful discovery window onto such light new degrees of freedom.  

\subsection{Naturalness}

The lightness of the Higgs boson is one of the most intriguing puzzles of modern particle physics. At the same time, null results in searches for electroweak-scale new physics have called the naturalness of the weak scale increasingly into question. In symmetry solutions to the hierarchy problem, the symmetry protecting the Higgs from sensitivity to high mass scales typically commutes with the Standard Model $SU(3)_c\times SU(2)_L\times U(1)_Y$ gauge group, giving rise to top partners that are charged under QCD. This guarantees a sizable LHC production cross-section for top partners with a mass at around the TeV scale, and therefore leads to tension with LHC searches that generically constrain top partners to have a TeV-scale mass. 
The most famous example of such theories is supersymmetry (SUSY). 

Some of these theories can be brought to agreement with LHC data at the cost of some tuning in the Higgs potential, as for example in models of split SUSY \cite{Arkani-Hamed:2004ymt,Giudice:2004tc,Arkani-Hamed:2004zhs} where scalar superpartner masses could be as high as the gauge unification scale with gaugino masses remaining around the electroweak scale.  Achieving such large splittings between scalars and gauginos requires, however, elaborate model-building \cite{Arkani-Hamed:2004ymt}.  SUSY models, in both their minimal and non-minimal realizations, have been a prime motivation for exotic Higgs decays. For example, gauge-mediated SUSY-breaking models can predict sizable branching ratios for $h\to\chi_2\chi_2,~\chi_2\to\chi_1\gamma$, where $\chi_2$ is a bino-like neutralino and $\chi_1$ a gravitino \cite{Mason:2009qh}.  Exotic Higgs decays have been especially well motivated in extensions of the minimal supersymmetric SM, such as the next-to-minimal supersymmetric SM with an approximate Peccei-Quinn symmetry, which predicts exotic Higgs decays such as $h\to\chi_1\chi_2,~\chi_2\to\chi_1 f\bar f$ where $f$ is a SM fermion and $\chi_2$ ($\chi_1$) is a bino-like (singlino-like) neutralino \cite{Draper:2010ew}.

A different approach to the hierarchy problem is to invoke symmetries that do not commute with the SM $SU(3)_c$ gauge group, leaving the weak scale to be stabilized by SM singlets or states with only EW quantum numbers. Examples of such theories are the Twin Higgs \cite{Chacko:2005pe} and Folded SUSY \cite{Burdman:2006tz} models, in which the top partners are singlets under the SM gauge group or have an EW charge, respectively, but carry charge under a mirror copy of the SM strong force.
In these ``Neutral Naturalness'' models, twin particle production proceeds through the Higgs portal operator that leads to a (generically small) mixing between the SM and twin Higgs sectors, thus allowing for both SM production of the heavier mirror Higgs as well as exotic Higgs decays into twin particles.\footnote{These models are specific realizations of the Hidden Valley scenario \cite{Strassler:2006im}.}
The resulting exotic Higgs branching fraction  depends on the mixing angle between the SM Higgs and  the heavier mirror Higgs, which is proportional to the ratio of SM to mirror Higgs vacuum expectation values (vevs).  The mirror Higgs vev in turn controls the scale of the twin particle masses and thus the residual tuning in the Higgs potential, so that bigger exotic branching fractions correspond to a less-tuned scalar sector.

The Twin Higgs model itself predicts invisible Higgs decays, as the produced twin particles ultimately cascade down to mirror photons and mirror neutrinos \cite{Burdman:2014zta}.  
The Fraternal Twin Higgs model adds mirror partners for only the subset of SM particles that contribute most strongly to the running of the Higgs mass \cite{Craig:2015pha}.  The lightest visibly-decaying partner particles in these theories are typically composite, dark glueballs and/or dark bottomonium \cite{Craig:2015pha}. 
Both the $0^{++}$ glueball $G_0$ and the scalar dark bottomonium state $\chi_{\tilde b 0}$ have the right quantum numbers to decay back to the SM through mixing with the Higgs.  Since these states are composite, their lifetimes depend very steeply on their masses.  In the better-understood case of the dark glueball, its lifetime can range from prompt or nearly prompt in the $\sim 40$ GeV regime, where direct two-body Higgs decays to pairs of $G_0$ are expected, to kilometers in the $\sim 10$ GeV regime, where a higher and variable multiplicity of particles is produced \cite{Craig:2015pha,Curtin:2015fna}.  In the high-multiplicity regime the typical signature of interest will thus be one displaced vertex plus missing energy, together with purely invisible decays.

A conceptually different way to address the hierarchy problem is realized by relaxion models \cite{Graham:2015cka}. In these models, the evolution of the Higgs mass in the early universe dynamically selects an electroweak scale that is parametrically smaller than the cutoff of the theory. This dynamical selection is realized thanks to a relaxion scalar field that scans the Higgs mass parameter from a large and positive cutoff energy down to negative values through a slow-roll potential. Once the Higgs mass parameter becomes negative, the Higgs gets a vev that modifies the Higgs-relaxion potential and eventually stops the rolling of the relaxion.
The relaxion generically stops its rolling at a point that breaks CP, leading to relaxion-Higgs mixing. The relaxion can be produced in exotic Higgs decays of the type $h\to\phi\phi$ with a branching ratio that depends on the specific implementation of the relaxion coupling to the Higgs
\cite{Flacke:2016szy}. Higgs decays provide a leading test of these relaxion models in the regime where relaxions are too heavy to be produced in meson decays \cite{Flacke:2016szy,Fuchs:2020cmm}. 

\subsection{Dark matter}

The most direct way for the Higgs to be connected to DM is for thermal freezeout to occur through exchange of the Higgs itself.  The Higgs portal operator readily allows interactions between the Higgs boson and a SM-singlet particle, of which the simplest and most minimal example is the renormalizable interaction between the SM Higgs and a scalar DM particle, 
$\Delta\mathcal{L} = -1/2\, \kappa S^2 |H| ^ 2$ \cite{Silveira:1985rk}.

In this minimal scalar model,
the Higgs-singlet quartic interaction $\kappa$
that yields the observed relic abundance is fixed as a function of the DM mass.\footnote{The singlet quartic self-coupling is a third parameter in this model, but for perturbative values of this coupling its cosmological impacts are limited \cite{Bernal:2015xba}.}  This coupling is generally large enough to yield unacceptably large Higgs invisible branching ratios when decays to DM are kinematically accessible.  In the region  $50\gev\lesssim m_{s}\lesssim m_h/2$, however, DM annihilations are resonantly enhanced in the early universe, allowing DM-Higgs couplings as small as $\mathcal{O}(10^{-4})$ to generate the correct relic abundance. 
 In this resonant region, DM-SM couplings are small enough that DM does not necessarily maintain kinetic equilibrium during freezeout, making DM annihilations less efficient \cite{Binder:2017rgn}.  Incorporating this effect in a careful calculation of the relic abundance increases the requisite value of $\kappa$  by as much as a factor of 2 compared to the prediction assuming kinetic equilibrium, depending on the modeling of scattering rates during the QCD phase transition.
 The current limit on the Higgs invisible width, $\br (h\to\inv) < 0.11$~\cite{ATLAS:2020kdi}, excludes the scalar model for $m_s<48$ GeV at 95\% CL.  
While invisible Higgs decays are very effective at excluding the low mass region, the constraints from direct detection experiments are even more stringent in the surviving resonant region (with standard modeling of the local DM halo). The spin-independent nuclear cross-sections generated by Higgs exchange give rise to appreciable signals in ton-scale liquid noble gas experiments.  Current results from XENON1T restrict the lower boundary of the resonant region to $m_{s}> 55.9 \, (57.2) $ GeV with minimal (maximal) kinetic decoupling effects \cite{XENON:2018voc}, leaving a viable albeit finely tuned window up to $m_h/2$.   Future liquid Xenon experiments have the potential to close the resonant window almost entirely \cite{DARWIN:2016hyl,PandaX:2018wtu}.  Meanwhile, the $s$-wave annihilation cross-section in this model makes indirect detection a competitive experimental test of the scalar singlet model, 
and in particular Fermi observations can be a powerful probe of the resonant regime \cite{Hardy:2018bph}. 
 
 The DM annihilation cross-section required to obtain the observed relic abundance can be reduced if the early universe underwent a period of early matter domination during or after DM freezeout \cite{Hardy:2018bph,Bernal:2018kcw,Chanda:2019xyl}. The smaller Higgs-singlet couplings necessary in these models allow the minimal scalar singlet model to survive both invisible Higgs decay constraints as well as direct detection exclusions in a broad mass range, depending on the timing and duration of the early matter-dominated era. 

The direct detection cross-section can also be suppressed by extending the singlet-Higgs sector.  Promoting the real scalar $S$ to a complex field $\Phi$, one can ensure that the stable DM particle is a pseudo-Goldstone boson of the (softly-broken) global $U(1)$ symmetry that takes $\Phi\to e^{i\alpha}\Phi$ \cite{Gross:2017dan}.  The  cross-section for elastic DM-SM particles is accordingly momentum-suppressed, and the leading contributions to the direct detection cross-section now appear at one loop, substantially below current experimental sensitivity.   Invisible Higgs decay modes, however, remain unsuppressed, and provide the most sensitive test of this model when they are kinematically accessible.  However, this enlarged parameter space comes with a price: the scalar sector in this theory has several additional free parameters 
and the invisible Higgs branching ratio is accordingly no longer uniquely predicted in terms of the DM mass and relic abundance.  The additional BSM Higgs boson $H_2$ in this theory provides further collider signatures, notably $H_2 \to \inv$, and can be an important source of constraints.  

Both fermionic and vector Higgs portal DM, described by the respective effective interactions  $\Delta\mathcal{L} = -( 1/\Lambda)( c_s\bar\chi\chi +  ic_p \bar \chi \gamma^5 \chi) |H| ^ 2$ and $-\lambda_V V_\mu V^\mu |H|^2$, are in some sense non-minimal models as  additional  degrees of freedom must be introduced  to construct even minimal UV completions of these Higgs portal interactions.  The surviving parameter space for both fermionic and vector models is qualitatively similar to that for the scalar model, with direct detection constraints restricting the DM mass to lie in the resonant regime $m_{DM}\lesssim m_h/2$ (or at TeV scales).  Collider and (to a lesser extent) direct detection signals however are sensitive to the additional fields required to UV-complete the Higgs portal interactions.

 Qualitatively distinct phenomenology arises in other classes of non-minimal models.
In one such class of models, DM freezes out by annihilating to SM final states through a BSM mediator, rather than the SM Higgs.  In Higgs portal models, this mediator is generally part of an extended Higgs sector, which inherits its coupling to the SM through mixing with (B)SM Higgs bosons.    A scalar singlet mediator, $S$, that inherits couplings to SM particles through Higgs mixing via the renormalizeable operator $S^2 |H|^2$ will generically lead to unacceptably large direct detection cross-sections\footnote{The major exception is the pseudo-Goldstone DM model discussed above, where the spontaneously broken global $U(1)$ ensures a cancellation between the contributions of the SM and BSM Higgs bosons in the direct detection cross-section.}.  A pseudo-scalar mediator, $a$, however, gives rise to a tree-level direct detection cross-section that is both spin-dependent and velocity-suppressed, making the one-loop spin-independent cross-section the leading source of direct detection constraints \cite{Ipek:2014gua}.  Coupling a pseudo-scalar singlet mediator to SM fermions in a UV-complete model requires either new (and generically stringently constrained) CP-violating couplings with the SM Higgs, or enlarging the SM Higgs sector to (e.g.) a two Higgs doublet model, which allows the mediator $a$ to mix with the doublet pseudo-scalar $A^0$.   In this case, exotic Higgs boson decays to pairs of mediators can easily be a leading probe of DM freezeout \cite{Ipek:2014gua}. The exotic branching ratio depends in detail on the properties of the heavy $A^0$ and is not uniquely predicted as a function of mediator mass.  The mediators can decay to either visible SM states or invisibly to DM, in principle allowing for invisible, visible, and semi-visible Higgs decays.  However, Fermi-LAT constraints on DM annihilations make it challenging to realize thermal relic DM with masses below 25 GeV in this model \cite{Fermi-LAT:2016uux,Fermi-LAT:2017opo},  thus substantially limiting the parameter space where both $h\to aa$ and $a\to \bar\chi\chi$ can  be realized.   
The parameter space of the 2HDM$+a$ model was recently surveyed in \cite{Robens:2021lov}.  

Adding BSM mediators to DM freezeout models also allows for the generic possibility that DM annihilations proceed directly to BSM mediator states, with no direct involvement of the SM.  
In models with such secluded annihilations the BSM mediator is generally lighter than DM and will often have a small coupling to the SM  that lets it decay into visible final states 
\cite{Pospelov:2007mp}. 
For these theories, the small mediator-SM coupling that determines both exotic Higgs decays and direct detection signals is parametrically decoupled from the annihilation cross-section that sets the DM relic abundance in the early universe, thereby ameliorating direct detection constraints.  Visible Higgs decays to mediator pairs are again a leading probe of this class of models. In the most minimal of such scenarios, the mediator is a single Higgs-mixed scalar \cite{Martin:2014sxa,Evans:2017kti}. In scenarios where a dark photon $Z_D$ mediates DM annihilations, the Higgs may also decay to $Z_D Z_D$ through mixing with the dark Higgs boson responsible for the mass of the $Z_D$ \cite{Bell:2016fqf}.

If the mediator decays dominantly through Higgs mixing, then the Higgs-mediator coupling cannot be arbitrarily small, as the mediator particle must be cosmologically short-lived in order not to ruin the successful predictions of BBN \cite{Fradette:2018hhl}.  Much more stringently, however, 
the assumption that DM shares the same temperature as the SM in the early universe relies on the mediator-Higgs coupling being large enough to keep the DM and the SM in kinetic equilibrium.  The ``thermalization floor'', i.e. the minimum value for the mediator-Higgs coupling that ensures secluded annihilations proceed while the DM is in equilibrium with the SM, represents an important experimental target \cite{Evans:2017kti}. For couplings below this value, the initial DM temperature must be separately specified from the SM temperature in order to obtain consistent predictions for the DM annihilation cross-section, which introduces dependence on additional model parameters.  Searches for $h\to ss$ at the LHC test mediator-Higgs couplings that are well above the thermalization floor \cite{Evans:2017kti}.

To summarize, $h\to $ invisible at levels at or below current limits can still be straightforwardly explained in simple models of thermal DM, though these theories are now either finely tuned or require non-minimal cosmological and/or particle content.   However, invisible decays are far from the only exotic decay mode motivated by DM model-building.   Simple theories of dark sector freezeout generically predict Higgs decays to pairs of new dark mediators, which typically result in (semi-)visible decays.  It is worth observing that  while the ``WIMP miracle'' ensures that DM freezing out via (extended) Higgs portal interactions 
has a mass scale not too dissimilar from the electroweak scale, a priori there is no compelling theoretical reason to prefer DM masses either above or below $m_h/2$. 
Detailed discussion of Higgs portal DM across the full viable mass range can be found in the recent reviews \cite{Arcadi:2019lka, Lebedev:2021xey}.

\subsection{Electroweak phase transition}

One long-standing motivation for introducing new states near the electroweak scale is to drive the electroweak phase transition (EWPT) strongly first-order.   
 Historically, electroweak baryogenesis has been a primary motivation for first-order EWPTs.  
 While constraints on BSM sources of CP violation have made electroweak baryogenesis model-building increasingly challenging, understanding the possible thermal histories of our universe remains an enduring and important question.  Strongly first-order EWPTs  produce stochastic gravitational wave signals, which may be seen in future gravitational wave detectors \cite{Caprini:2019egz}.
 The electroweak phase transition in the SM is a crossover \cite{Kajantie:1996mn,Csikor:1998eu}, but new degrees of freedom interacting with the Higgs boson can generate first-order EWPTs through either loop effects or tree-level couplings with the SM. 
 In models that feature new particles with $m< m_h/2$, the most salient possibility is 
 SM singlet scalars, which can affect the EWPT at tree level.   
After the discovery of the SM Higgs boson, 
most work on collider signatures of singlet-catalyzed strongly first-order EWPTs has focused on the regime where Higgs decays to exotic states are kinematically forbidden, as the vast majority of parameter space with open BSM decay modes yields exotic branching ratios that are now experimentally unacceptable \cite{Profumo:2007wc,Ghorbani:2018yfr}.  However, there is still a narrow region of viable parameter space where a light SM-singlet scalar can drive the electroweak phase transition strongly first-order, yielding exotic Higgs decays into either visible \cite{Kozaczuk:2019pet,Carena:2019une} or invisible \cite{Kozaczuk:2019pet} final states. The branching fraction for such exotic decays is bounded from below, as the singlet coupling to the SM Higgs boson cannot be arbitrarily small and still successfully drive the phase transition first order.  The HL-LHC will be able to test the lower bound on the visible exotic branching fraction consistent with a strongly first-order EWPT for scalars with mass $m\gtrsim 20 $ GeV, while at lower scalar masses lepton colliders will be needed to conclusively probe the lower bound \cite{Kozaczuk:2019pet}. 

\section{Benchmark Scenarios}
\label{sec:benchmarks}

Searches for visible exotic Higgs decays are often carried out in the context of phenomenological models that guide overall analysis strategies.  The best-studied of these models  consider the minimal scenarios where the Higgs decays to pairs of singlet states, $h\to ss$, $aa$, or $Z_D Z_D$.  Here $s$ denotes a CP-even scalar, $a$ a CP-odd scalar, and $Z_D$ a vector boson.  
The key properties of these particles for exotic decay searches are governed by the operator(s) (or ``portal'') through which they interact with the SM, which determine both their production rate in Higgs decays as well as their decays to SM particles. In particular, the choice of decay portal controls the dominant final states.

Here we discuss three of the most common benchmark scenarios: (i) SM$+$s, where $s$ is a Higgs-mixed scalar boson; (ii) SM$+$ALP; 
(iii) SM$+$v, where the vector, $Z_D$, kinetically mixes with the SM hypercharge gauge boson.
While these benchmark scenarios certainly do not lead to all the possible exotic decays of interest, they capture the key phenomenology predicted in a wide variety of theories, motivate searches in a broad range of final states, and provide a useful guide to assessing the reach and motivation of various experimental analysis strategies.

\subsection{SM+s}
\label{sec:benchmark_sm_plus_s}

One of the simplest models leading to exotic Higgs boson decays occurs when a new real scalar is added to the SM, mixing with the Higgs boson and inheriting its couplings to SM fields.  This model is intimately connected to many of the theoretical models discussed above: the scalar can be motivated by electroweak phase transitions, relaxion models, neutral naturalness (in which case the scalar is typically long-lived), or it could be a mediator in a dark matter model. The minimal Lagrangian describing the system is given by
\begin{equation}\label{eq:SH}
\mathcal L = \mathcal L_{{\rm{kin}}} + \frac{\mu_s ^ 2}{2} S^2-\frac {\lambda_s} {4!} S ^ 4-\frac{\kappa}{2}S ^ 2 |H | ^ 2 + \mu ^ 2 |H |^ 2-\lambda |H | ^ 4,
\end{equation}
where we have imposed a discrete symmetry taking $S\to -S$. Depending on the parameters of the scalar potential, both $S$ and $H$ can get non-zero vevs, $v_s$ and $v_h$ respectively:
\begin{equation}
v_s^2 =\, \frac{6 \left(2\lambda\mu_s^2 - \kappa\mu^2 \right)}{2\lambda\lambda_s-3\kappa^2},\,\phantom{space} v_h^2 =\, \frac{2 \lambda_s \mu^2 - 6 \kappa\mu_s^2}{2\lambda\lambda_s-3\kappa^2}\,.
\end{equation}
In this case, the two scalar mass eigenstates, $h$ and $s$ with masses $m_h$ and $m_s$ respectively, are related to the gauge eigenstates by the mixing angle
\begin{equation}\label{eq:theta}
\tan \theta = \frac{ \kappa v_h v_s}{\lambda v_h^2 - \frac16 \lambda_s v_s^2+\sqrt{\left( \lambda v_h^2 - \frac16 \lambda_s v_s^2 \right)^2 +  \kappa^2 v_h^2 v_s^2 }}\simeq \frac{ \kappa v_h v_s}{ m_{h}^2-m_s^2}\,,
\end{equation}
where the latter relation holds only when $m_s$ is quite different than $m_h$ and $\kappa$ is small. 
For small values of $\theta$ (as motivated by LEP constraints \cite{Robens:2015gla}), the branching ratio for $h\to ss$ depends on BSM parameters only through the mixed quartic $\kappa$ and on the phase space available for the decay,
\begin{equation}
Br(h\to ss)\simeq\frac{v^2\kappa^2}{32\pi m_h\Gamma_h}\sqrt{1-\frac{4m_s^2}{m_h^2}},
\end{equation}
where $\Gamma_h$ is the total Higgs  width. 

The scalar $s$ decays preferably to the heaviest SM particles that are kinematically accessible.  The $s$ lifetime is proportional to $\tan^2\theta$, while its branching fractions are independent of  $\theta$. 
To evaluate the branching ratios of $s$ to SM states as a function of its mass, we use the results of \cite{Gershtein:2020mwi} for $m_s<20$ GeV and as given by the LHC Higgs Cross Section Working Group~\cite{LHCHiggsCrossSectionWorkingGroup:2016ypw} at larger values of the mass. 
The predictions at lower masses include decays to hadrons, particularly pions and kaons, following the treatment of \cite{Winkler:2018qyg}, while at higher masses contributions from heavier particles such as $W$ and $Z$ are also considered. The two calculations differ at the $\sim 30\%$ level for hadronic decay modes at 20~GeV. It would be desirable to have a consistent treatment across the full mass range in the future.
The experimental results discussed in Sec.~\ref{sec:exp_prompt_sm_plus_s} are independent of $\theta$ as long as the mixing angle is large enough to allow $s$ to decay promptly, while the LLP searches considered in Section~\ref{sec:exp_llp_sm_plus_s} are sensitive to the $s$ lifetime and thus $\theta$.

\subsubsection{Scalars vs.~pseudoscalars} 

Collider searches for exotic Higgs decays to spin-zero states are largely insensitive to the CP properties of $s$, and, thus, searches for the $h\to ss$ signatures predicted by the SM+s benchmark also have similar sensitivity to pseudoscalars, $a$,
as arising, for example, in theories where light singlet pseudoscalars mix with pseudoscalars belonging to a doublet representation of $SU(2)$.  Thus the benchmark SM+s is a useful reference model to characterize experimental sensitivity to a broader class of models featuring a new light state with Yukawa-weighted decays.

However, the couplings of $a$ to SM fermions and gauge bosons, and thus the branching ratios of $a$, depend on the parameters of the doublet state it mixes with, such as $m_A$ and $\tan \beta$ in two Higgs doublet models (see e.g. \cite{Dolan:2014ska,Haisch:2018kqx}). Moreover, in the GeV-scale regime where the new particle can decay directly to hadrons, its hadronic matrix elements will depend on its CP properties.
For this reason, $a$ and $s$ branching ratios in the sub-GeV mass range
will be substantially different. 

A new pseudoscalar that mixes with a 2HDM doublet pseudoscalar can also give rise to exotic Higgs decays $h\to Za$, which cannot be realized in the SM+s model.  The $h\to Za$ decay mode would probe parameter regions away from the alignment limit~\cite{Craig:2013hca}.

\subsection{SM+ALP}
\label{sec:benchmark_sm_plus_alp}

Another interesting class of exotic Higgs boson decays can arise when a new ALP, $a$, is added to the SM.  Such ALPs appear in well-motivated extensions of the Standard Model, e.g. as a way to address the strong CP problem, as mediators between the SM and a hidden sector (through the so-called ``Axion Portal''), or simply as pseudo-Nambu-Goldstone bosons in extensions of the SM with a broken global symmetry.  
Models with light ALPs generically lead to exotic Higgs decays into ALPs.

The ALP can couple to the Higgs through the dimension-six and seven operators 
\begin{equation}\label{ALPEFT}
\mathcal L\supset \frac{C_h}{\Lambda^2}(\partial_\mu a)(\partial^\mu a)H^\dagger H+\frac{C_Z}{\Lambda^3}(\partial_\mu a)(H^\dagger i D_\mu H +h.c.)H^\dagger H\,.
\end{equation}
Note that at dimension five there are no $a$ couplings to the Higgs doublet.\footnote{Here we have assumed that electroweak symmetry is broken linearly. Nonlinear EWSB can allow for additional $h Z a$ interactions, making $h\to aZ$ a leading probe of such theories \cite{Brivio:2017ije}.} After electroweak symmetry breaking, these two operators will lead to the decays $h\to aa$ \cite{Dobrescu:2000jt,Dobrescu:2000yn} and $h\to aZ$ \cite{Bauer:2017nlg}, respectively. The corresponding widths are given by
\begin{eqnarray}
  \Gamma(h\to aa) 
  &=& \frac{v^2 m_h^3}{32\pi\Lambda^4} \left|C_h\right|^2
    \left( 1 - \frac{2m_a^2}{m_h^2} \right)^2 \sqrt{1 - \frac{4m_a^2}{m_h^2}} \,,\\
    \Gamma(h\to Za) &=& \frac{m_h^3v^4}{64\pi\Lambda^6} \left| C_Z\right|^2 
     \lambda^{3/2}\bigg(\frac{m_Z^2}{m_h^2},\frac{m_a^2}{m_h^2}\bigg) \,,
\end{eqnarray}
where $\lambda(x,y)=(1-x-y)^2-4xy$.

Once produced, the ALP can decay back to the SM thanks to the dimension-five effective Lagrangian (following the notation of \cite{Bauer:2017ris})
\begin{equation}\label{Leff}
\begin{aligned}
 {\cal L}_{\rm eff}
  &=  \frac{\partial^\mu a}{\Lambda} \sum_F\,\bar\psi_F\,C_F\,\gamma_\mu\,\psi_F
+ g_s^2\,C_{GG}\,\frac{a}{\Lambda}\,G_{\mu\nu}^A\,\tilde G^{\mu\nu,A} \\[-1mm]   
& \quad\mbox{}
    + g^2\,C_{WW}\,\frac{a}{\Lambda}\,W_{\mu\nu}^A\,\tilde W^{\mu\nu,A}
    + g^{\prime\,2}\,C_{BB}\,\frac{a}{\Lambda}\,B_{\mu\nu}\,\tilde B^{\mu\nu} \,,
    \end{aligned}
\end{equation}
where $G_{\mu\nu}^A$, $W_{\mu\nu}^A$ and $B_{\mu\nu}$ are the field strength tensors of $SU(3)_c$, $SU(2)_L$ and $U(1)_Y$, and $g_s$, $g$ and $g'$ denote the corresponding coupling constants. The sum in the first line extends over the chiral fermion multiplets $F$ of the SM. The quantities $C_F$ are hermitian matrices in generation space. The couplings in Equation~\ref{Leff} will also induce loop corrections to the $h\to aa$ and $h\to Za$ rates.

Depending on the values of these couplings, the ALP can decay to a large set of final states, including $a\to \bar f f,~a\to\gamma\gamma,~a\to gg$.  When the fermionic operators dominate the ALP decays, the resulting phenomenology is generically a realization of the Yukawa-weighted models discussed in the previous subsection.  Generically, however, and in contrast to light scalar models, theories containing derivatively-coupled ALPs  predict substantial branching ratios of the ALP into gluons and/or photons. 
This makes the exotic decays $h\to aa\to\gamma\gamma jj$, $h\to aa\to\gamma\gamma\gamma\gamma$, $h\to aa\to jjjj$ particularly relevant to test ALP theories. 
When $m_a\lesssim 10$ GeV, its decay products  begin to appear collimated in the detector. For photonically-decaying ALPs, the resulting photon-jet-like signature in the detector can present an interesting reconstruction task \cite{Sheff:2020jyw}, and can in some parts of parameter space potentially contribute to searches for the SM $h\to \gamma\gamma$ decay \cite{Draper:2012xt,Ellis:2012zp}. 

In understanding the range of signatures described by the effective field theory of Equations~\ref{ALPEFT} and~\ref{Leff}, it can be useful to keep in mind concrete UV completions, which make definite and model-dependent predictions for the relative sizes of the Wilson coefficients as well as the scale $\Lambda$. 
A variety of UV completions 
have been studied in the literature, including some that realize viable lepton flavor-violating scenarios with $h\to aa$ followed by $a\to \tau \ell$ \cite{Davoudiasl:2017zws,Davoudiasl:2021haa} (similar signatures were considered in \cite{Evans:2019xer}). In such scenarios $a$ can be either prompt or long-lived, leading to interesting displaced flavor-violating signatures.

\subsection{SM+v}
\label{sec:benchmark_sm_plus_v}

Dark photon theories feature a gauge boson from a spontaneously broken $U(1)_D$ gauge symmetry that kinetically mixes with the SM hypercharge gauge boson, $\hat B$.  The minimal dark photon model has become a standard benchmark for feebly-interacting particle searches, and is a  common building block in theories of DM. 
The relevant gauge terms in the Lagrangian are 
\begin{equation}
\mathcal{L}\supset -\frac{1}{4}\hat Z^{\prime\mu\nu}\hat Z^\prime_{\mu\nu}+\frac{\epsilon}{2\cos\theta}\hat B^{\mu\nu}\hat Z^\prime_{\mu\nu}+\frac{1}{2}m_{\hat Z^\prime}^2\hat Z^{\prime\mu}\hat Z^{\prime}_\mu\,,
\label{eq:darkphotonL}
\end{equation}
where $\theta$ is the Weinberg angle, and $\epsilon$ is the kinetic mixing parameter. After EWSB, the mass eigenstates $Z$ (corresponding to the SM $Z$ boson) and $Z_D$ will have non-zero $\hat Z$ and $\hat B$ components.
The dark photon mass term in Equation~\ref{eq:darkphotonL} can come either from a dark Higgs, $S$, or from the Stueckelberg mechanism. In the former case, the Lagrangian will also contain the interactions of the Higgs with the scalar as in Equation~\ref{eq:SH}, as well as the interaction of the scalar with two dark photons.

If the dark photon is light enough, the kinetic mixing leads to the exotic decay $h\to ZZ_D$ with a width given by
\begin{eqnarray}
\Gamma(h\to Z Z_D) &=& \frac{\eta ^2 \, \sin^2\theta \, m_Z^2 \,m_{Z_D}^2}{16 \,\pi \, v^2 \,m_h^3 \,\left(m_Z^2-m_{Z_D}^2\right)^2} \left(-2 m_{Z_D}^2 \left(m_h^2-5 m_Z^2\right)+m_{Z_D}^4+\left(m_h^2-m_Z^2\right)^2\right)  \nonumber 
\\
&&\times \, \sqrt{-2 m_h^2 \left(m_{Z_D}^2+m_Z^2\right)+\left(m_Z^2-m_{Z_D}^2\right)^2+m_h^4}\,,
\label{eq:htoZDZ}
\end{eqnarray}
where we have defined $\eta\equiv \epsilon/(\cos\theta\sqrt{1-\epsilon^2/\cos^2\theta})$. 
The kinetic mixing also leads to the decay $h\to Z_DZ_D$, but the width of this decay is suppressed by $\epsilon^4$, making it highly sub-leading. 
However, if the $Z_D$ mass is generated by a dark Higgs mechanism, the dark scalar-Higgs mixing induced by the Lagrangian in Equation~\ref{eq:SH} also contributes to the decay $h\to Z_DZ_D$,
\begin{eqnarray}\label{eq:htoZDZD}
\Gamma(h \rightarrow Z_D Z_D) &=& {\kappa'}^2 \ \frac{1}{32 \pi}  \ \frac{v^2}{m_h} \sqrt{1 - \frac{4 m_{Z_D}^2}{m_h^2}}  \  \frac{(m_h^2 + 2 m_{Z_D}^2)^2 - 8 (m_h^2 - m_{Z_D}^2)m_{Z_D}^2}{m_h^4},
\end{eqnarray}
where we have have introduced the dimensionless parameter $\kappa'\equiv\kappa \,{m_h^2}/|m_h^2-m_s^2|$,
with $\kappa$ the quartic interaction of the dark scalar and the Higgs, $\kappa |S|^2|H|^2$.

Through kinetic mixing, the dark photon acquires couplings to the SM fermions:
\begin{equation}
\mathcal{L}\supset  g_L Z_{D\mu}\bar f_L\gamma^\mu f_L \: + \: g_R Z_{D\mu}\bar f_R\gamma^\mu f_R \: \,,
\end{equation}
where the couplings are given by
\begin{align}\label{eq:Zpcoups}
	& g_L =\frac{g_W}{\cos \theta}\left (-\sin \alpha \: (\cos^2 \theta_W  \: T_3 - \sin^2 \theta \:  Y_L) + \cos \alpha \: \eta \: \sin \theta_W \: Y_L\right)\simeq e Q \epsilon, \nonumber \\
	& g_R= \frac{g_W}{\cos \theta} \left(-\sin \alpha \: (-\sin^2 \theta  \: Y_R ) + \cos \alpha \: \eta \: \sin \theta \: Y_R\right)\simeq e Q \epsilon.
\end{align}
Here $\alpha$ is the mixing angle between $Z$ and $Z_D$, and is proportional to the kinetic mixing parameter, $\alpha\propto\epsilon$. $T_3$ and $Y$ are, respectively, the third component of the isospin and the hypercharge of the SM fermion, where we use the convention $Q = T_3+Y$, with $Q$ the fermion
electric charge. The last equalities are only valid in the limits $\epsilon\ll 1$ and $m_{Z_D}\ll m_Z$.

Because of these couplings, the dark photon will decay to SM fermions with partial widths given by 
\begin{equation}\label{eq:width-ZDff}
\Gamma(Z_D \to \bar f f) = \frac{N_c}{24 \pi  m_{Z_D}} \sqrt{1-\frac{4 m_f^2}{m_{Z_D}^2}} \left(m_{Z_D}^2 \left(g_L^2+g_R^2\right)-m_f^2 \left(-6 g_L g_R+g_L^2+g_R^2\right)\right),
\end{equation}
proportional to $\epsilon^2$. This tree-level formula is a good approximation for $m_{Z_D}$ above the $b\bar b$ threshold. Below this threshold, one should include experimental information and higher order QCD calculations (see e.g.~\cite{Curtin:2014cca}). Overall, in the mass range of interest, the dark photon has a large probability to decay to SM leptons, $Br(Z_D\to\ell^+\ell^-)\sim 30-50\%$, where we are summing over electrons and muons. The branching ratios of $Z_D$ do not depend on $\epsilon$, while the $Z_D$ lifetime is inversely proportional to $m_{Z_D}\epsilon^2$.

Our focus in Secs.~\ref{sec:exp_prompt_sm_plus_v}, \ref{sec:exp_llp_sm_plus_v} will be on the minimal scenario where the exotic decay of primary interest is $h\to Z_D Z_D$, where the branching ratio $Br(h\to Z_D Z_D)$ is controlled by the mixing between the SM and the dark Higgs \cite{Schabinger:2005ei,Gopalakrishna:2008dv}. Depending on the mass splittings in the theory, the decay $h\to ss\to 4 Z_D$ may also be available, resulting in very high multiplicity final states \cite{Izaguirre:2018atq}.  
In the minimal dark photon model, direct searches for  $h\to Z Z_D $ with $\mathcal O(100\,{\mathrm{fb}}^{-1})$ data can start to test regions of the $(\epsilon, m_{Z_D})$ parameter space that are not excluded by precision electroweak constraints on $\epsilon$ \cite{Curtin:2014cca}, making these searches the leading probe of the minimal model for masses $m_{Z_D}\gtrsim 10$ GeV.
Non-minimal models that allow for additional mass mixing between $Z$ and $Z_D$ can predict larger branching fractions for this exotic decay than expected from kinetic mixing alone \cite{Davoudiasl:2013aya,Lu:2017uur}. 

Another non-minimal variant of the dark photon model adds a dark neutralino and higgsino to give the Higgsed dark $U(1)$ theory a supersymmetric matter content \cite{Falkowski:2010cm,Falkowski:2010gv,Chan:2011aa}.  This scenario, which we refer to as a SUSY dark photon model, can yield semi-invisible and potentially high-multiplicity decays, depending on the details of the hidden sector spectrum.

\FloatBarrier

\section{Experimental Status}
\label{sec:exp}

Exotic Higgs boson decays to new particles have very rich phenomenology featuring many possible final states, mass regimes, and lifetime ranges. Several searches have been performed at the Tevatron and LHC for such signals. Typically, the searches assume that the observed Higgs boson at 125~GeV is produced in accordance with SM expectations but decays to new particles. This is because new particles introduced in the models described in Section~\ref{sec:benchmarks} have a subleading effect on the Higgs boson production cross-section, since their couplings are expected to be much smaller than the electroweak and top couplings to the Higgs~\cite{LHCHiggsCrossSectionWorkingGroup:2016ypw}. The searches target different Higgs boson production modes, mainly due to trigger considerations, as well as different mass and lifetime ranges, often depending on the object identification and reconstruction techniques available, as well as the analysis strategy pursued. Experimental searches can be broadly categorized into those targeting scenarios where the new particle decays promptly to SM particles (Section~\ref{sec:exp_prompt}), and where the new particle is long-lived, resulting in a displaced or invisible decay (Section~\ref{sec:exp_llp}). 

The total Higgs boson width cannot be measured in a model-independent way at the LHC. However, a global fit of Higgs measurements involving SM final states can constrain the branching ratio into exotic final states under the mild assumption that BSM physics does not enhance the Higgs widths to $W$ and $Z$ bosons. Recent analyses set an upper limit at 95\% CL of about 16\% ~\cite{ATLAS-CONF-2021-053} 
on the ``undetected'' Higgs branching fraction (i.e., final states that do not contribute to searches for SM or invisible decays). 
Given the assumptions necessary to extract this limit, and the finite sensitivity expected even after the completion of the Run 2 and Run 3 analyses, there is ample room for interestingly large exotic branching fractions while remaining compatible with current measurements. This is very strong motivation for pursuing the complementary strategy of performing direct searches. 

Higgs boson decay products carry momenta ranging from a few GeV to few tens of GeV, which can be challenging for triggers at hadron colliders. Searches for final states with electrons, muons, taus, or photons typically target Higgs bosons produced by gluon-fusion (ggF), which has the largest cross section. In hadronic final states, the decay products are typically too soft to satisfy trigger requirements, and searches rely on additional forward jets, targeting Higgs boson production via vector boson fusion (VBF), or on additional leptons from $W$ or $Z$ boson decays produced in association with the Higgs boson ($Wh/Zh$). 

Further challenges arise at the reconstruction and identification level. Since the decay products are often soft and may overlap in the detector, standard algorithms can reject events or objects produced in exotic Higgs decays. Dedicated identification and reconstruction techniques have been developed for several signatures, targeting low $p_{T}$ or overlapping decay products, such as $b$-jets~\cite{ATLAS:2020ahi}, taus~\cite{CMS:2019spf}, and muons~\cite{CMS:2018jid}. Such dedicated algorithms extend the sensitivity of searches to cover new scenarios. 

Searches for long-lived particles (LLPs), which decay with measurable displacements in the detector, offer a great opportunity to detect new physics because of the striking signatures. These searches also present unique challenges for triggering, reconstruction, and identification and often require dedicated techniques.  LLP signatures can resemble noise, pileup, or mis-reconstructed objects in the detector, which are often rare and hard to model reliably in simulation. Therefore, dedicated analysis methods to estimate backgrounds are also often needed.

\subsection{Summary of prompt searches}
\label{sec:exp_prompt}

Searches for exotic Higgs boson decays to new particles that subsequently decay promptly to SM particles have been performed in several final states and mass ranges. Current searches primarily target Higgs boson decays to intermediate on-shell neutral particles, 
$h \rightarrow ss/aa/vv \rightarrow XX~YY$ and $h \rightarrow Za/Zv \rightarrow \ell\ell~XX$
where $s(a)$ is a new (pseudo)scalar, $v$ a new vector, and $X$ and $Y$ are SM particles that are pair-produced in each new particle decay: electrons, muons, taus, photons, b-jets, or hadronic jets. Higgs decays to a pair of new particles decaying to different final states are also explored ($X \neq Y$). Table~\ref{tab:prompt_summary} summarizes experimental results for prompt decays to SM particles.

\begin{table}[htbp]
\tabcolsep7.5pt
\caption{Summary of the latest prompt searches for $h\rightarrow ss/aa/vv$ or $Za/Zv$, including merged (m) and resolved (r) final states. 
The interpretations included in each result are listed, following the categorization presented in Section~\ref{sec:benchmarks}, and $m$ denotes the new particle mass.  The two mass ranges of the last listed search correspond to the interpretation SM+s and SM+v, respectively. }
\label{tab:prompt_summary}
\begin{center}
\begin{tabular}{@{}p{0.4cm}m{0.65cm}|@{\hskip 0.1cm}c@{\hskip 0.1cm}|@{\hskip 0.1cm}c@{\hskip 0.1cm}|@{\hskip 0.1cm}c@{\hskip 0.1cm}|@{\hskip 0.1cm}c@{\hskip 0.1cm}|@{\hskip 0.1cm}c@{\hskip 0.1cm}|@{\hskip 0.1cm}c@{}}
\hline
\hline
\multicolumn{2}{l|}{Decay} & Mode & Reference & \makecell{$\sqrt{s}$\\(TeV)} & \makecell{$\int \mathcal{L}$\\(fb$^{-1}$)} & $m$ (GeV) & Interpretations\\
\hline
\hline
\multicolumn{8}{c}{\boldmath $h\rightarrow ss/aa/vv$}\\
\hline
\hline
$eeee$  &(r)   & ggF & CMS \cite{CMS-PAS-HIG-19-007}   & 13 & 137 & 4-8, 11.5-62.5 & SM+v, SM+ALP \\
        & (r) & ggF & ATLAS \cite{ATLAS-CONF-2021-034} & 13 & 139 & 15-60 & SM+s, SM+v \\
\hline
$ee\mu\mu$  & (r)   & ggF & CMS \cite{CMS-PAS-HIG-19-007}   & 13 & 137 & 4-8, 11.5-62.5 & SM+v, SM+ALP \\
            & (r)   & ggF & ATLAS \cite{ATLAS-CONF-2021-034} & 13 & 139 & 15-60 & SM+v \\
\hline
$\mu\mu\mu\mu$  & (m)   & ggF & D0 \cite{Abazov_2009} & 1.96 & 4.2 & 0.2143-3 & SM+s, SM+v \\
                & (r)   & ggF & CMS \cite{CMS:2018jid} & 13 & 35.9 & 0.25-8.5 & SM+s, dark SUSY  \\
                & (r)   & ggF & CMS \cite{CMS-PAS-HIG-19-007}   & 13 & 137 & 4-8, 11.5-60 & SM+v, SM+ALP \\
                & (m/r) & ggF & ATLAS \cite{ATLAS-CONF-2021-034} & 13 & 139 & 1.2-2, 4.4-8, 12-60 & SM+s, SM+v \\
\hline
$\mu\mu\tau\tau$    & (m/r)    & ggF & D0 \cite{Abazov_2009} & 1.96 & 4.2 & 3.6-19 & SM+s \\
                    & (m/r) &  ggF & ATLAS \cite{ATLAS:2015unc} & 8 & 20.3 & 3.7-50 & SM+s \\
                    & (m/r) &  ggF & CMS \cite{CMS:2020ffa}  & 13 & 35.9 & 3.6-21 & SM+s \\
                    & (r)   &  ggF & CMS  \cite{CMS:2018qvj} & 13 & 35.9 & 15-62.5 & SM+s \\
\hline
$\tau\tau\tau\tau$  & (m)   & ggF & CMS \cite{CMS:2019spf} & 13 & 35.9 & 4-15 & SM+s \\
\hline
$bb\mu\mu$  & (r)   & ggF & ATLAS \cite{ATLAS:2021hbr} & 13 & 139 & 18-60 & SM+s \\ 
            & (r)   & ggF & CMS \cite{CMS:2018nsh} & 13 & 35.9 & 20-62.5 & SM+s \\
\hline
$bb\tau\tau$& (r)   & ggF & CMS \cite{CMS:2018zvv} & 13 & 35.9 & 15-60 & SM+s \\
\hline
$bbbb$    & (m)   & $Zh$ & ATLAS \cite{ATLAS:2018pvw} & 13 & 36.1 & 15-30 & SM+s \\ 
        & (r)   & $Wh/Zh$ & ATLAS \cite{ATLAS:2020ahi} & 13 & 36.1 & 20-60 & SM+s \\  
\hline
$\gamma\gamma\gamma\gamma$& (r) & ggF & ATLAS \cite{ATLAS:2015rsn} & 8 & 20.3 &10-62 & SM+s \\
                          & (r) & ggF & CMS  \cite{CMS-PAS-HIG-21-003} & 13 & 132 & 15-60 & SM+s \\               
\hline
$\gamma\gamma gg$   & (r)        & VBF & ATLAS \cite{ATLAS:2018jnf} & 13 & 36.7 & 20-60 & SM+s \\
\hline
\hline
\multicolumn{8}{c}{\boldmath $h\rightarrow Za/Zv$}\\
\hline
\hline
$gg$ & (m)    & ggF & ATLAS \cite{ATLAS:2020pcy} & 13 & 139 & 0.5-4 & SM+s \\
$ss$ & (m)    & ggF & ATLAS \cite{ATLAS:2020pcy} & 13 & 139 & 1.5-3 & SM+s \\
\hline 
$ee$ & (r)  & ggF & CMS \cite{CMS-PAS-HIG-19-007}   & 13 & 137 & 4-8, 11.5-35 & SM+v\\
     & (r)  & ggF & ATLAS \cite{ATLAS-CONF-2021-034} & 13 & 139 & 15-55 & SM+v \\
\hline
$\mu\mu$ & (r)  & ggF & CMS \cite{CMS-PAS-HIG-19-007}   & 13 & 137 & 4-8, 11.5-35 & SM+v\\
         & (r)  & ggF & ATLAS \cite{ATLAS-CONF-2021-034} & 13 & 139 & 15-30/15-55
         & SM+s, SM+v \\
\hline
\hline
\end{tabular}
\end{center}
\end{table}


\subsubsection{Branching Ratio and Mass Sensitivity}
\label{sec:exp_br_mass}

\begin{figure}[htbp]
\includegraphics[width=\linewidth]{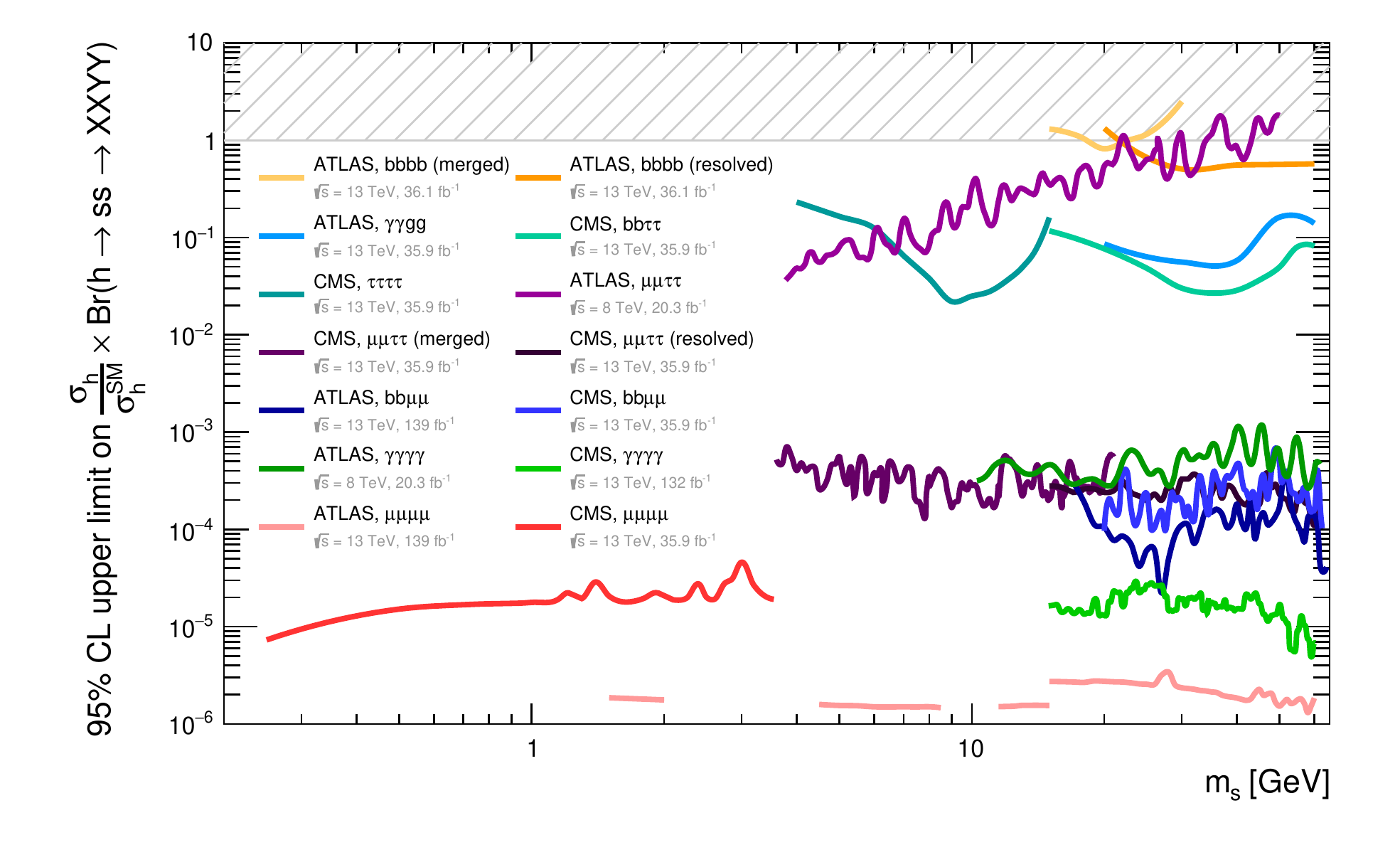}
\caption{Observed 95\% CL upper limits on $\sigma_h/\sigma_h^{SM} Br(h\rightarrow ss \rightarrow XX YY)$ where $s$ is a new scalar decaying to pairs of SM particles $X$ and $Y$, and $\sigma_h^{SM}$ is the SM Higgs boson production cross section. The most recent $h\rightarrow ss$ analyses from Table~\ref{tab:prompt_summary} are included.}
\label{fig:ExpSummary} 
\end{figure}

The sensitivity of current searches for Higgs decays to a pair of new scalars is summarized in Figure~\ref{fig:ExpSummary}, which shows the 95\% CL upper limits on the branching ratio excluded by each final state as a function of the mass of the new particle. The limits on $\sigma/\sigma_h^{SM}\times Br(h \rightarrow ss \rightarrow XX~YY)$ do not depend on the (model-dependent) branching ratios of the decays of the new scalar, and make it possible to compare the overall sensitivity for different channels and mass ranges. None of the searches pursued so far are able to distinguish between odd and even parity for the new scalar, and hence the symbols $s$ and $a$ may be used interchangeably. 
Decays to new vector particles  
can result in significantly different acceptance, so dedicated interpretations are needed. 

Current searches cover a mass range for the new particle  from $m \gtrsim 2m_{\mu}\approx 0.2$~GeV to $m_h/2 \approx 62.5$~GeV. The lower mass range is exclusively covered by the $\mu\mu\mu\mu$ decay channel, which spans the full mass range except for challenging regions close to the $J/\psi$ and $\Upsilon$ mass peaks in some analyses. These searches currently set the strongest constraints on the branching ratio for a given exclusive final state, down to $\sim 10^{-6}$~\cite{ATLAS-CONF-2021-034,CMS:2018jid}. Searches 
for the $\mu\mu\tau\tau$ and $\mu\mu bb$ decays are currently able to place limits on exclusive branching ratios down to $\sim 10^{-5}-10^{-4}$ in the mass range from $m \gtrsim  2m_{\tau}\approx 4$~GeV to $m_h/2 \approx 62.5$~GeV~\cite{Abazov_2009,ATLAS:2015unc,CMS:2020ffa,CMS:2018qvj,ATLAS:2021hbr,CMS:2018nsh}.
Searches in the $bb\tau\tau$~\cite{CMS:2018zvv}, $\tau\tau\tau\tau$~\cite{CMS:2019spf}, and $bbbb$~\cite{ATLAS:2018pvw,ATLAS:2020ahi} final states reach sensitivity in the range $\sim 10^{-2}-10^{-1}$ for the mass range $m \gtrsim 2m_{b}\sim 10$~GeV. Searches sensitive to photons also place significant constraints on the branching ratio, including limits down to $10^{-5}$ for $m \approx 12-62.5$~GeV in the $\gamma\gamma\gamma\gamma$ final state~\cite{ATLAS:2015rsn,CMS-PAS-HIG-21-003} and $10^{-1}$ for $m\approx 20-60$ for $\gamma\gamma jj$~\cite{ATLAS:2018jnf}.

There are currently fewer experimental searches targeting decay modes to a new pseudoscalar or vector boson produced with a $Z$ boson. 
A search for hadronic scalar decays~\cite{ATLAS:2020pcy} sets upper limits on $Br(h \rightarrow Z a \rightarrow \ell\ell XX)$ as low as 0.35 
for $m_a$ in the range $0.5-4$~GeV, assuming $a$ decays to either gluon pairs or strange or charm quark pairs. Decays to muons and electrons are also explored~\cite{ATLAS-CONF-2021-034}, setting limits on the branching ratio down to $\approx 5\times 10^{-5}$ in the $15-30$~GeV or $15-55$~GeV mass range for the pseudoscalar or vector cases, respectively.

The current reach of all the searches in Table~\ref{tab:prompt_summary} is limited by statistics, so updated analyses using all available data will improve the sensitivity. More sophisticated analyses, including new reconstruction and identification techniques, can help complete the coverage of the full mass range. Additional searches in uncovered channels may also bring additional sensitivity and are interesting cross checks in case an excess is observed. 

\subsubsection{SM+s}
\label{sec:exp_prompt_sm_plus_s}

Searches for decays to a new light scalar, $s$, often focus on the heaviest particles that are kinematically allowed in the scalar decay. 
Decays to muons are considered for $m \gtrsim 2m_{\mu}\approx 0.22$~GeV and are particularly important in the lowest mass range until decays to taus may also become important, $m \gtrsim 2m_{\tau}\approx 3.6$~GeV. Finally in the mass range $m \gtrsim 2m_{b}\approx 8.4$~GeV, several searches also target decays to $b$-jets. 

\begin{figure}[htbp]
\includegraphics[width=\linewidth]{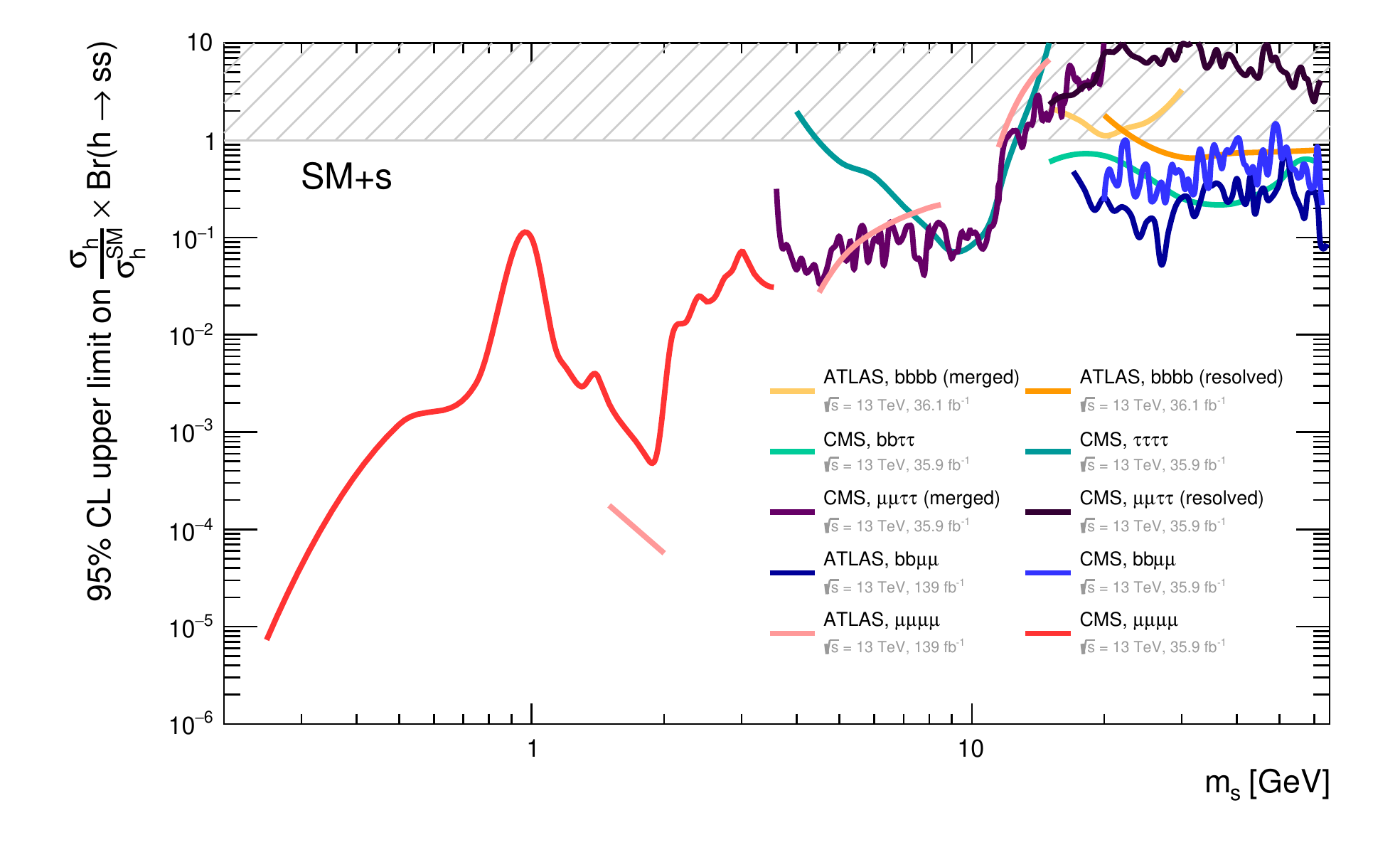}
\caption{Observed 95\% CL upper limits on $\sigma_h/\sigma_h^{SM} Br(h\rightarrow ss)$ in the SM+s scenario where $s$ is a new Higgs-mixed scalar, from a selection of the most recent analyses in Table~\ref{tab:prompt_summary}. The branching fractions of the new scalar to SM particles are taken from~\cite{Gershtein:2020mwi,LHCHiggsCrossSectionWorkingGroup:2016ypw}, as described in Section~\ref{sec:benchmark_sm_plus_s}.
\label{fig:SM_S}}
\end{figure}

Figure~\ref{fig:SM_S} shows the upper limits on $Br(h\to ss)$ in the SM+s scenario, using the branching ratios for the new scalar predicted by the minimal model of Sec. \ref{sec:benchmark_sm_plus_s}. 
The strongest constraints appear at the lowest masses from the $\mu\mu\mu\mu$ mode, setting branching ratio limits down to 10$^{-5}$. Between the $J/\psi$ and the $\Upsilon$ thresholds, the sensitivity steadily decreases to about 10$^{-1}$. The sensitivity from the $\mu\mu\mu\mu$ channel is comparable to the $\mu\mu\tau\tau$ and $\tau\tau\tau\tau$ channels in the $\sim (7-10)$ GeV mass range. In the highest mass range $m \gtrsim 2m_b$, several final states including $b$-quarks ($bbbb$, $bb\mu\mu$ and $bb\tau\tau$) constrain branching ratios down to about $10^{-1}$.
A combination of results from different channels, especially in the intermediate and higher mass range where several decay modes have comparable sensitivity, would extend the reach beyond the current limits.

\subsubsection{SM+ALP}
\label{sec:exp_prompt_sm_plus_alp}

There are fewer searches interpreted in the context of SM+ALP scenario compared to the SM+s and SM+v cases, as summarized in Table~\ref{tab:prompt_summary}.  As an illustrative example, the benchmark scenario with couplings to fermions $C_F=1$ and to gauge bosons $C_{BB}=C_{WW}=C_{GG}=1/(4\pi)^2$ discussed in \cite{Bauer:2017ris} has appreciable fermionic branching ratios, but enhanced branching ratios to photons and gluons compared to the SM+s model. 
The searches listed in Table~\ref{tab:prompt_summary} can be reinterpreted in this particular benchmark SM+ALP model 
and are able to exclude $\sigma / \sigma_h^{SM} \times Br(h \rightarrow aa)$ as small as $10^{-1}$. The enhanced gluon branching fraction reduces current sensitivity to this benchmark point compared to SM+s, as the reach is mostly driven by searches relying on fermionic final states.  
Current searches for $gg\gamma\gamma$ and $\gamma\gamma\gamma\gamma$ have lower sensitivity to probe $Br(h\to aa)$ for this specific benchmark point, but they can be very important in other benchmarks  where $C_F\ll 1$.
There is a strong case to expand the current experimental program to cover more final states involving photons and gluons.

\subsubsection{SM+v}
\label{sec:exp_prompt_sm_plus_v}

\begin{figure}[!htb]
 \centering
       \includegraphics[width=0.47\textwidth]{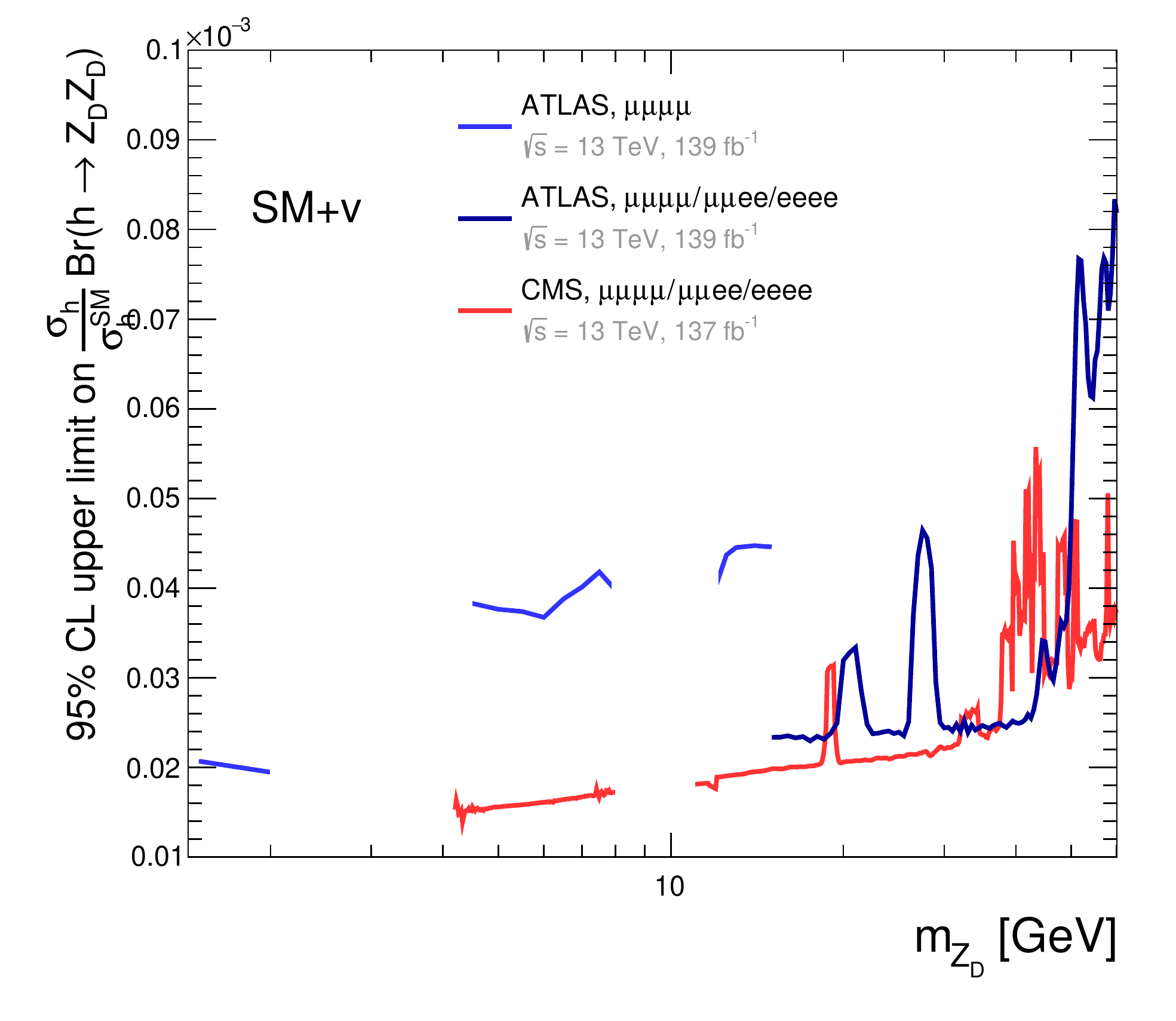}
       \includegraphics[width=0.47\textwidth]{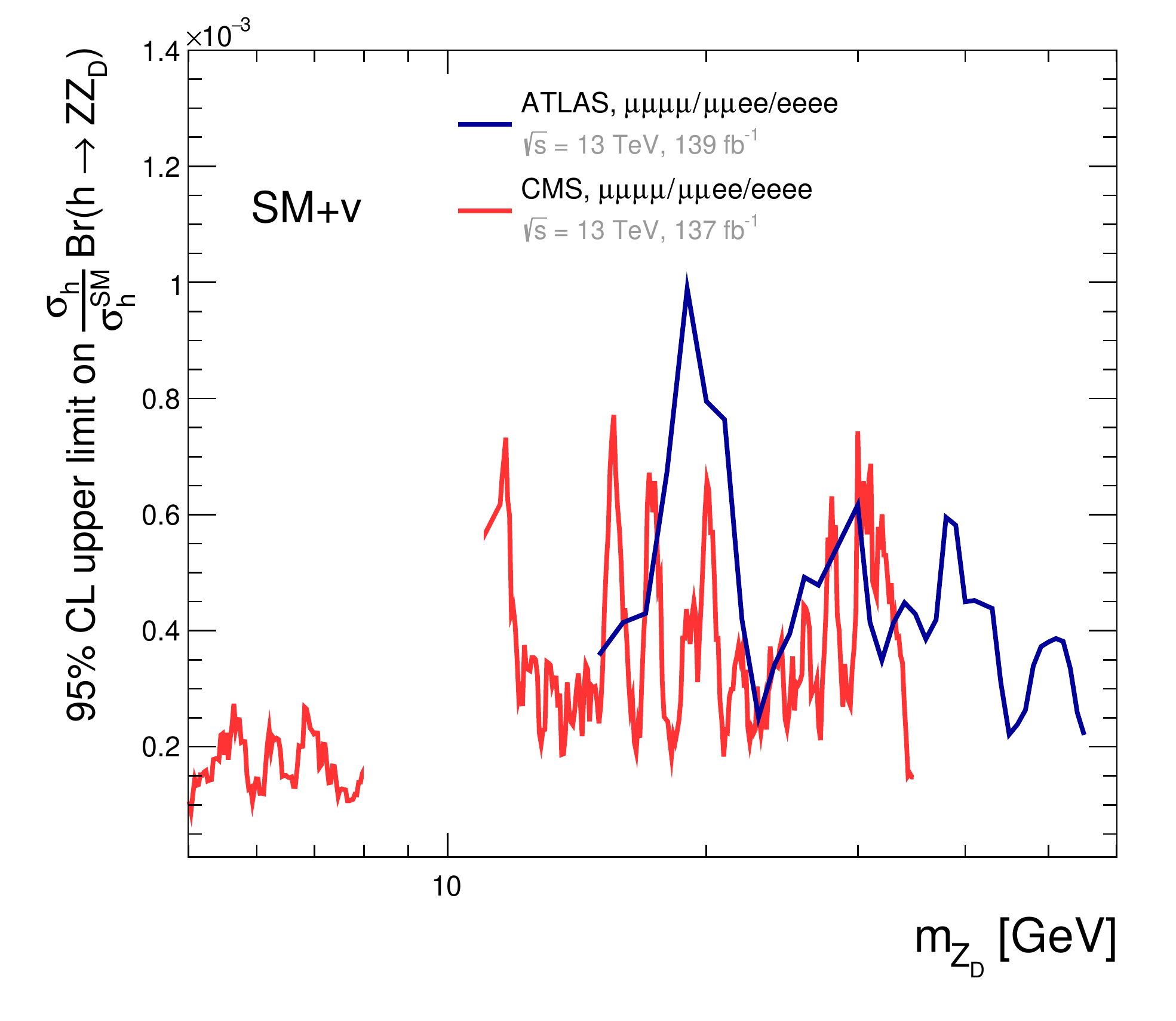}
\caption{Observed 95\% CL upper limits on (left) $\sigma_h/\sigma_h^{SM} Br(h\rightarrow Z_{D}Z_{D})$ and (right) $\sigma_h/\sigma_h^{SM} Br(h\rightarrow ZZ_{D})$ in the SM+v scenario, where $Z_{D}$ is a kinetically-mixed dark vector, from the  
most recent searches in Table~\ref{tab:prompt_summary}. 
The ATLAS~\cite{ATLAS-CONF-2021-034} limits correspond to a low mass $4\mu$ search (light blue, only available for the $h\rightarrow Z_{D}Z_{D}$ interpretation) and a higher mass search that considers muon and electron decays (dark blue). The CMS search \cite{CMS-PAS-HIG-19-007} includes muon and electron decays in the full mass range for both $h\rightarrow Z_{D}Z_{D}$ and $h\rightarrow Z Z_{D}$. The branching fractions of the new vector to SM particles follow the prescriptions in~\cite{Curtin:2014cca}, as described in Section~\ref{sec:benchmark_sm_plus_v}. 
}
\label{fig:SM_V}
\end{figure}

New light vectors arising in models with kinetic mixing are expected to have sizeable branching ratios to leptons, as described in Section~\ref{sec:benchmark_sm_plus_v}. Figure~\ref{fig:SM_V} shows the latest upper limits on the exotic Higgs branching ratio to final states with a dark $Z$ boson, $Z_D$, from different searches with lepton final states after considering the $Z_D\to f\bar{f}$ decay width given in Equation~\ref{eq:width-ZDff}. The figure summarizes the limits obtained for both $\sigma_h / \sigma_h^{SM} Br(h \rightarrow Z_D Z_D)$ (left panel) and $\sigma / \sigma_h^{SM} \times Br(h \rightarrow Z Z_D)$ (right panel).
The CMS search \cite{CMS-PAS-HIG-19-007} looks for $h \rightarrow Z_D Z_D$ in the mass ranges 4.2-8 and 11.5-60 GeV, and for $h \rightarrow Z Z_D$ in the mass ranges 4.2-8 and 11.5-35 GeV, exploiting in both cases $Z_D$ decays to electrons or muons in the full mass range. The ATLAS search \cite{ATLAS:2018coo} considers $Z_D$ decays to electrons or muons in the higher mass range ($h \rightarrow Z_D Z_D$: 15-60 GeV,  $h \rightarrow Z Z_D$: 15-55 GeV), with similar sensitivity to CMS. In the lower mass range (4.4-8 GeV, 12-15 GeV, and also below the $J/\psi$ mass, 1.2-2 GeV), the same ATLAS search is restricted to $h\rightarrow Z_D Z_D$ and considers only the $4\mu$ final state (see light blue line in the left panel of the figure). 
Both ATLAS and CMS interpret the results for $h\to Z_DZ_D$ as limits on the coupling constant $\kappa$ that controls the mixing between the dark scalar and the Higgs, and for $h\to ZZ_D$ as limits on $\epsilon$ that controls the mixing between the dark vector and SM hypercharge  (see Section~\ref{sec:benchmark_sm_plus_v}).

Available searches can be improved by including electrons in the lower mass range, by performing dedicated background estimates in the $J/\psi$ and $\Upsilon$ pole regions, and by extending the search below $m_{Z_D}<1$~GeV, using techniques similar to the ones in the CMS search~\cite{CMS:2018jid}. 
Other searches for SM+v include displaced decays, and are discussed next. 

\FloatBarrier

\subsection{Summary of displaced and invisible signatures}
\label{sec:exp_llp}

Experiments at the LHC have performed several searches for Higgs boson decays to both visibly-decaying LLPs 
and invisible BSM particles. Dedicated searches targeting LLPs that travel macroscopic distances in the detector rely on the reconstruction and identification of displaced objects. Even though they were not originally designed to specifically detect LLPs, ATLAS and CMS have sensitivity to a variety of LLP signals. 
Searches have been performed in both hadronic and leptonic channels. LHCb, covering the pseudorapidities $2<\eta<5$,
is designed to observe displaced decays and has also performed such searches.

In the case of Higgs boson decays to LLPs, available searches cover a broad range of masses and lifetimes $c\tau$. 
The LLPs are assumed to be on-shell and neutral, giving rise to processes of the type  
$h \rightarrow ss/vv \rightarrow XX~YY$, where $s$ is a new scalar, $v$ a new vector and $XX$ and $YY$ are pairs of SM particles from each new boson decay.  In several cases the LLP signature is very distinctive and the searches have low backgrounds, so the analyses require reconstruction of only one of the two decays. These searches are referred to as $h \rightarrow s/v + X \rightarrow YY + X$, where $X$ refers to all possible decays of the second particle in the pair. 
Table~\ref{tab:llp_summary} summarizes the latest searches for exotic Higgs decays to LLPs. 

\begin{table}[htbp]
\tabcolsep7.5pt
\caption{Summary of the latest LLP searches for $h\rightarrow ss/vv$. $m$ and $c\tau$ denote the new particle mass and lifetime, respectively. 
}
\label{tab:llp_summary}
\begin{center}
\begin{tabular}{@{}l@{\hskip 0.1cm}|@{\hskip 0.1cm}c@{\hskip 0.1cm}|@{\hskip 0.1cm}c@{\hskip 0.1cm}|@{\hskip 0.1cm}c@{\hskip 0.1cm}|@{\hskip 0.1cm}c@{\hskip 0.1cm}|@{\hskip 0.1cm}c@{\hskip 0.1cm}|@{\hskip 0.1cm}c@{\hskip 0.1cm}|@{\hskip 0.1cm}c@{}}
\hline
\hline
Decay & Mode & Reference & Method & {\small $\sqrt{s}$} (TeV) & $\int \mathcal{L}$ (fb$^{-1}$) & $m$ (GeV) & $c\tau$ (m) \\
\hline \hline
\multicolumn{7}{c}{{\bf{SM+s: \boldmath $h\rightarrow ss$ or \boldmath $ s + X$, $s$ long-lived}}}\\ \hline
\hline
$bbbb$    & $Wh/Zh$ & ATLAS \cite{ATLAS:2020ahi} & {\small prompt reinterp.} & 13 & 36.1 & 20-60  & $10^{-4} - 10^{-2}$ \\ \hline

$bbbb$ &  \multirow{3}{*}{ggF} & \multirow{3}{*}{LHCb \cite{LHCb:2017xxn}} & \multirow{3}{*}{{\small disp. jets}} & \multirow{3}{*}{7,8} & \multirow{3}{*}{2.0} & \multirow{3}{*}{25-50} 
    & \multirow{3}{*}{$10^{-3}-10^{-1}$}  \\ 
$cccc$ &  & & & & &  \\ 
$ssss$ &  & & & & & \\  \hline 

$bbbb$    & \multirow{2}{*}{$Zh$} & \multirow{2}{*}{CMS \cite{CMS:2021uxj}} & \multirow{2}{*}{{\small Z+disp. jets}} & \multirow{2}{*}{13} & \multirow{2}{*}{117} & \multirow{2}{*}{15-55} 
& \multirow{2}{*}{$10^{-3} - 1$}\\ 
$dddd$    &                     & &  & &  &\\ \hline 

$bbbb$ & $Zh$ & ATLAS \cite{ATLAS:2021jig} & {\small Z+disp. jets} & 13 & 139 & 16-55 
& $10^{-3} - 1$ \\ \hline 

$bbbb$ & \multirow{2}{*}{ggF} & \multirow{2}{*}{CMS \cite{CMS:2020iwv}} & \multirow{2}{*}{{\small disp. jets}} & \multirow{2}{*}{13} & \multirow{2}{*}{132} & \multirow{2}{*}{15-55} 
& \multirow{2}{*}{$10^{-3} - 10$} \\
$dddd$ &  & &  &  & & & \\ \hline 

$bbbb$ & \multirow{3}{*}{ggF} & \multirow{3}{*}{ATLAS \cite{ATLAS:2019qrr}} & \multirow{3}{*}{{\small CalRatio}} & \multirow{3}{*}{13} & \multirow{3}{*}{10.8, 33.0} & \multirow{3}{*}{5-55} 
& \multirow{3}{*}{$10^{-1} - 10^{3}$}\\
$cccc$               &    & & & & & \\
$\tau\tau\tau\tau$ &    &  & & & & \\ \hline

$bbbb$            & \multirow{3}{*}{ggF} & \multirow{3}{*}{ATLAS \cite{ATLAS:2019jcm}} &  \multirow{3}{*}{{\small ID+MS DVs}} & \multirow{3}{*}{13} & \multirow{3}{*}{33.0} & \multirow{3}{*}{8-55}  & \multirow{3}{*}{$10^{-1} - 10$} \\ 
$cccc$  &  & & & & & \\
$\tau\tau\tau\tau$      &  & & & & & \\ \hline 

$bbbb$    & \multirow{3}{*}{ggF} & \multirow{3}{*}{CMS \cite{CMS:2021juv}} & \multirow{3}{*}{{\small hadronic MS}} & \multirow{3}{*}{13} & \multirow{3}{*}{137} &  14-55 & \multirow{3}{*}{$10^{-1} - 10^{4}$} \\ 
$dddd$               &    & & & & & 7-55  \\
$\tau\tau\tau\tau$ &    &  & & & & 7-55 \\ \hline

$bbbb$ & \multirow{3}{*}{ggF} & \multirow{3}{*}{ATLAS \cite{ATLAS:2018tup}} &  \multirow{3}{*}{{\small MS1+MS2 DV}} & \multirow{3}{*}{13} & \multirow{3}{*}{36.1} & \multirow{3}{*}{5-40} 
& \multirow{3}{*}{$10^{-1} - 10^{3}$}\\
$cccc$               &    & & & & &  \\
$\tau\tau\tau\tau$ &    &  & & &  & \\ \hline

$bbbb$ & \multirow{3}{*}{ggF} & \multirow{3}{*}{ATLAS \cite{ATLAS-CONF-2021-032}} & \multirow{3}{*}{{\small MS2 DV}} & \multirow{3}{*}{13} & \multirow{3}{*}{139} & \multirow{3}{*}{5-55} 
& \multirow{3}{*}{$10^{-1} - 10^{2}$}\\
$cccc$               &    & & & &  & \\
$\tau\tau\tau\tau$ &    &  & & &  & \\ \hline

$e\mu$+X            & \multirow{3}{*}{ggF} & \multirow{3}{*}{CMS \cite{CMS:2021kdm}} & \multirow{3}{*}{{\small disp. leptons}} & \multirow{3}{*}{13} & \multirow{3}{*}{113-118} & \multirow{3}{*}{30-50}  & \multirow{3}{*}{$10^{-3} - 10^{1}$} \\ 
$\mu\mu$+X  &  & & & & & \\
$ee$+X      &  & & & & & \\ \hline 
\hline 
\multicolumn{7}{c}{{\bf{SM+v: \boldmath $h\rightarrow Z_{D} Z_{D}$ or \boldmath $Z_{D} + X$, $Z_{D}$ long-lived}}}\\ \hline
\hline 

$\mu\mu\mu\mu$  & \multirow{2}{*}{ggF} & \multirow{2}{*}{CMS \cite{CMS:2021ogd}} & \multirow{2}{*}{{\small dimuon scouting}} & \multirow{2}{*}{13} & \multirow{2}{*}{101} & \multirow{2}{*}{0.5-50} & \multirow{2}{*}{$10^{-4} - 10$} \\
$\mu\mu$+X &    & & & & & \\ \hline

$\mu\mu$+X  & ggF & ATLAS \cite{ATLAS:2018rjc} & {\small MS dimuon} & 13 & 32.9 & 20-60 & $10^{-3} - 10^{2}$\\ \hline 
\hline 
\multicolumn{7}{c}{{\bf{SM+v: dark SUSY long-lived}}}\\ \hline
\hline 

$\mu\mu$+X  & \multirow{2}{*}{ggF} & \multirow{2}{*}{ATLAS \cite{ATLAS:2019tkk}} & \multirow{2}{*}{{\small disp. lepton-jets}} & \multirow{2}{*}{13} & \multirow{2}{*}{36.1} & \multirow{2}{*}{$0.2-3.6$} & \multirow{2}{*}{$10^{-3} - 1$}\\ 
$hh$+X &    & & & & & \\ \hline
$hh$+X & ggF  &  ATLAS~\cite{ATL-PHYS-PUB-2020-007} & {\small recast of \cite{ATLAS:2019qrr}} & 13 & 10.8, 33.0 & $\sim 0.4$ & $10^{-3} - 10^{-1}$ \\ \hline
$\mu\mu\mu\mu$  & ggF & CMS \cite{CMS:2018jid}  & {\small disp. muons}  & 13 & 35.9 & 0.25-8.5 & 0-1  \\
\hline\hline
\end{tabular}
\end{center}
\end{table}

Existing searches cover a large number of final states including large impact-parameter tracks and displaced vertices (DVs) reconstructed in the inner tracking detector (ID), displaced hadronic jets or photons in the calorimeter, and displaced signals from jets or leptons in the muon system (MS).  Dedicated searches based on DV reconstruction using tracks in the ID matched to a ``displaced jet'' achieve sensitivity to LLPs with lifetimes in the range from a few mm to about a meter. These searches often rely on the ability to reconstruct tracks with large impact parameter~\cite{ATLAS:2017zsd}. Given the difficulty triggering on signals in the ID, these searches typically rely on other accompanying objects, such as leptons from the decays of a $Z$ boson produced in association with the Higgs boson~\cite{ATLAS:2020ahi,CMS:2021uxj}. Searches targeting signatures in the calorimeter and MS are sensitive to the lifetime range from a few cm to hundreds of meters~\cite{ATLAS:2019qrr,CMS:2021juv}. Calorimeter and muon signals can be used in the trigger, in some cases with dedicated strategies for displaced decays, and such searches therefore target Higgs bosons produced by gluon fusion. Examples include the ``CalRatio'' trigger that identifies trackless displaced jets in the ATLAS hadronic calorimeter by searching for signals with low electromagnetic fraction compared to the hadronic fraction~\cite{ATLAS:2019qrr} and the ``MS vertex'' trigger that searches for clusters of particles in the ATLAS MS~\cite{ATLAS:2018tup, ATLAS-CONF-2021-032}. Dedicated reconstruction algorithms are also used to reconstruct displaced decays using signals in the muon detectors, including multi-track vertices in the ATLAS MS which are used to perform searches with one (MS1) and two (MS2) vertices~\cite{ATLAS:2018tup, ATLAS-CONF-2021-032}, and clusters of hits in the CMS endcap MS~\cite{CMS:2021juv} which are used to reconstruct hadronic and electromagnetic showers (hadronic MS). Combinations of signatures also offer sensitivity to several signals, such as the ATLAS search for both DVs in the ID and the MS~\cite{ATLAS:2019jcm}. The CMS search for displaced leptons~\cite{CMS:2021kdm} targets transverse impact parameters between 0.01~cm and 10~cm and is thus sensitive to LLP lifetimes from 1~mm to 10~m. Searches targeting longer lifetimes benefit from the distinctiveness of the signal and the few SM backgrounds. For the longest lifetimes, the sensitivity is limited by the size of the detector. 

One of the main challenges to search for low mass LLPs is that these signals have large SM backgrounds and are therefore difficult to trigger on. One limitation of the trigger systems is the bandwidth for offline processing and data storage. New strategies such as ``scouting'', where the event size is reduced as a trade-off for recording larger rates of events, make it possible to target signals such as displaced leptons with small displacements and low masses, extending the sensitivity beyond the reach of traditional searches~\cite{CMS:2021ogd}. Complementary searches targeting the largest lifetimes rely on muon reconstruction in the MS alone, without a corresponding signal in the inner detectors and can achieve sensitivity with a single vertex due to the distinctive, low-background signal~\cite{ATLAS:2018rjc}. This strategy uses dedicated triggers to identify tracks in the muon detectors without requiring matches to an ID track or pairs of collimated muons in the MS. 

\FloatBarrier

LLPs may be invisible if they decay outside the detector volume or if the standard algorithms are unable to reconstruct the signature. Therefore searches for invisible decays ($h\rightarrow  \mathrm{E}_{T}^{\mathrm{miss}}$), in which the decay product escapes detection or is undetectable, and semi-invisible decays ($h\rightarrow s/v + \mathrm{E}_{T}^{\mathrm{miss}}$), where one of the new particles decays to SM particles and the other is invisible, are also sensitive to exotic Higgs boson decays to LLPs. Available searches for these signals are summarized in Table~\ref{tab:hinv_summary}. Both invisible and semi-visible searches rely on the presence of substantial missing transverse energy, $\mathrm{E}_{T}^{\mathrm{miss}}$, to identify the Higgs decay. Current experimental bounds on invisible Higgs boson decays are mainly from searches for Higgs bosons produced by VBF~\cite{CMS:2019ajt,ATLAS-CONF-2020-052} or in association with a $Z$ boson~\cite{ATLAS-CONF-2021-029,CMS:2020ulv}. Although the combination of searches from ATLAS and CMS using the full Run 2 dataset has not been completed, branching ratios greater than 0.11 are already excluded. All current upper limits in each production mode can be found in Table~\ref{tab:hinv_summary}.  

\begin{table}[htbp]
\tabcolsep7.5pt
\caption{Summary of the latest searches for invisible and semi-visible Higgs boson decays, including available interpretations for semi-visible final states and observed (expected) upper limits (UL) at 95\% CL for invisible signatures. 
} 
\label{tab:hinv_summary}
\begin{center}
\begin{tabular}{@{}l|@{\hskip 0.1cm}c@{\hskip 0.1cm}|@{\hskip 0.1cm}c@{\hskip 0.1cm}|@{\hskip 0.1cm}c@{\hskip 0.1cm}|@{\hskip 0.1cm}c@{\hskip 0.1cm}|c@{}}
\hline
\hline
\multicolumn{6}{c}{\boldmath $h\rightarrow s/v + \mathrm{E}_{T}^{\mathrm{miss}}$}\\ \hline 
Decay & Mode & Reference & $\sqrt{s}$ (TeV) & $\int \mathcal{L}$ (fb$^{-1}$) & Interpretations\\
\hline
$\mathrm{E}_{T}^{\mathrm{miss}}$ + $\gamma$ & VBF & CMS \cite{CMS:2020krr}   & 13 & 130 & SM+v \\ %
                          & VBF & ATLAS \cite{ATLAS:2021pdg} & 13 & 139 & SM+v \\ 
                          & $Zh$  & CMS \cite{CMS:2019ajt}   & 13 & 137 & SM+v  \\ 
                          & ggF, $Zh$ & CMS \cite{CMS:2015ifd} & 8 & 19.4 & Other \\ \hline 
$\mathrm{E}_{T}^{\mathrm{miss}} +  bb$      & $Zh$ & ATLAS \cite{ATLAS:2021edm} & 13 & 139  & NMSSM  \\ \hline           
\hline
\hline
\multicolumn{6}{c}{\boldmath $h\rightarrow \mathrm{E}_{T}^{\mathrm{miss}}$}\\
\hline
Decay & Mode & Reference & $\sqrt{s}$ (TeV) & $\int \mathcal{L}$ (fb$^{-1}$) & Br(H$\rightarrow$Inv) UL\\ \hline 
$\mathrm{E}_{T}^{\mathrm{miss}}$   & VBF & ATLAS \cite{ATLAS-CONF-2020-008}  & 13 & 139 &  0.13 (0.13) \\   
                 & VBF & CMS \cite{CMS-PAS-HIG-20-003} & 13 & 138 &  0.17 (0.11)   \\  
                 & $Z(ll)h$  & ATLAS \cite{ATLAS-CONF-2021-029} & 13 & 139 &  0.18 (0.18)    \\ 
                 & $Z(ll)h$ & CMS \cite{CMS:2020ulv} & 13 & 137 & 0.29 (0.25) \\ 
                 & ggF & ATLAS \cite{ATLAS:2021kxv} & 13 & 139 & 0.34 (0.39)\\
                 & ggF, $V(qq)h$ & CMS \cite{CMS:2021far} & 13 & 137 & 0.278 (0.253) \\ 
                 & $tth$ & ATLAS \cite{ATLAS-CONF-2020-052}& 13 & 139 & 0.40 (0.36)   \\ 
                 & $tth$ & CMS \cite{CMS:2019bke}& 13 & 35.9 & 0.46 (0.48)  \\ 
                 & Combination & ATLAS \cite{ATLAS-CONF-2020-052} & 7, 8, 13 & 4.7+20.3+139 & 0.11 (0.11) \\
                 & Combination & CMS \cite{CMS:2018yfx} & 7, 8, 13 & 4.9+19.7+38.2 &  0.19 (0.15)  \\ 
\hline
\hline
\end{tabular}
\end{center}
\end{table}

In several scenarios, LLP searches have sensitivities that are greater than those for promptly decaying particles. Searches at lower lifetimes and masses typically lose sensitivity due to the presence of larger backgrounds. The trigger and identification of low mass LLP signatures can also be challenging since the decay products may overlap and the ability to reconstruct DVs significantly degrades for nearby particles whose trajectories are parallel. For higher masses, the decay products tend to have larger opening angles and in some cases the objects become so non-pointing that the reconstruction and identification efficiency may degrade. The LHC already has sensitivity to a range of LLP signals and this capability can be extended by exploiting the large datasets and benefiting from trigger and detector improvements in the future (see Section~\ref{sec:future}).

The following sections describe hadronic searches with an interpretation focusing on exotic Higgs decays to a pair of new scalars (Section~\ref{sec:exp_llp_sm_plus_s}) and leptonic searches focusing on new vectors (Section~\ref{sec:exp_llp_sm_plus_v}). Studies of the differences in acceptance for these scenarios have not been performed for LLP signatures. There are no ATLAS or CMS interpretations of searches for exotic Higgs decays to long-lived ALPs. Searches for displaced photons would be of particular interest for scenarios with ALPs. 

\FloatBarrier

\subsubsection{SM+s}
\label{sec:exp_llp_sm_plus_s}

\begin{figure}[htb]
    \centering
        \includegraphics[width=\textwidth]{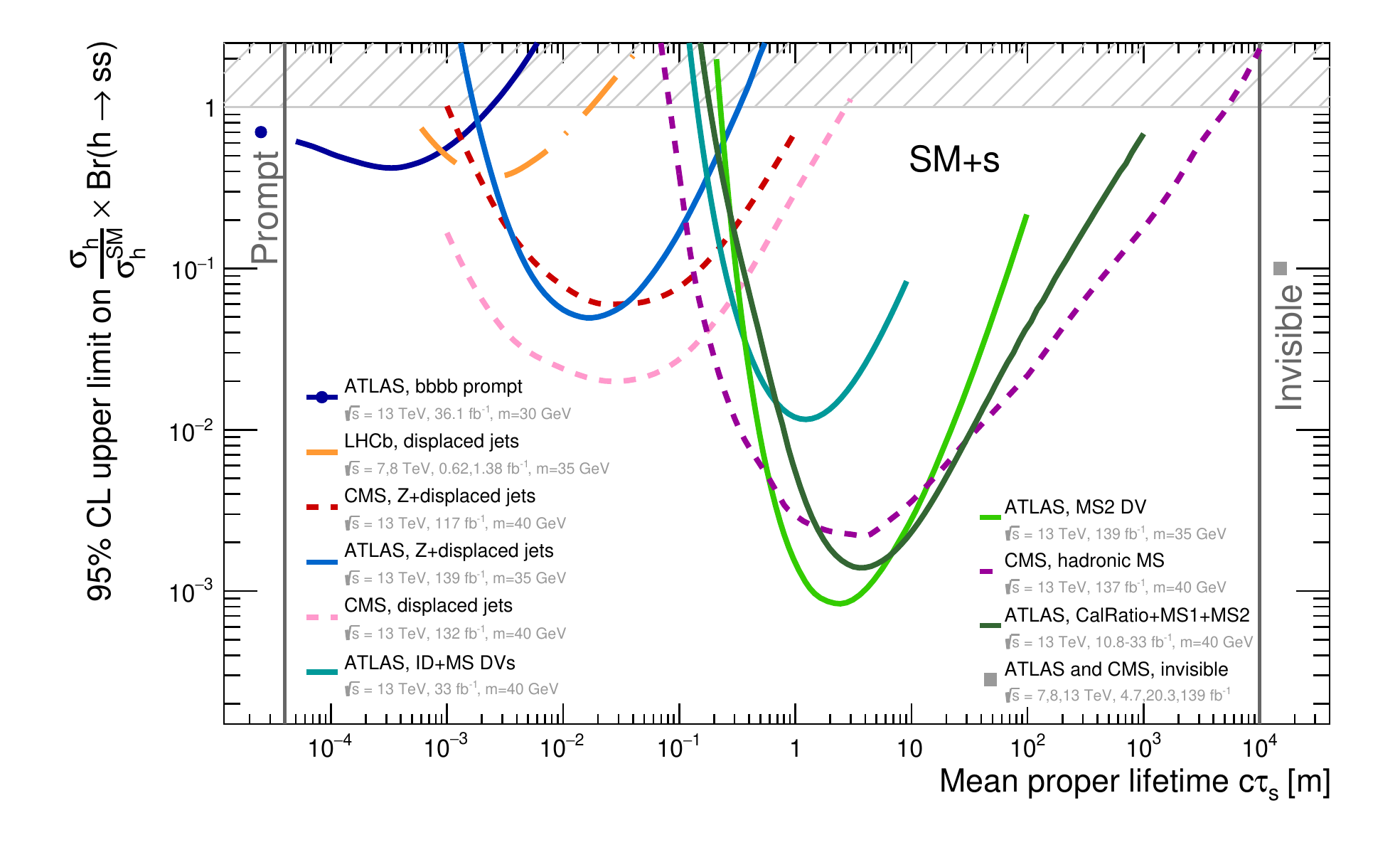}
\caption{Observed 95\% CL upper limits on $\sigma_h/\sigma_h^{SM} Br(h\rightarrow ss)$ in the SM+s scenario where $s$ is a new long-lived scalar decaying hadronically. The analyses are summarized in Table~\ref{tab:llp_summary} and interpreted according to the procedure described in the text.
The branching fractions of the new scalar boson to SM particles follow the prescriptions in~\cite{LHCHiggsCrossSectionWorkingGroup:2016ypw}, as described in Section~\ref{sec:benchmark_sm_plus_s}. The current best limit on invisible Higgs boson decays (see Table~\ref{tab:hinv_summary}) is shown for comparison (gray marker on the right). The bounds from the reinterpretation of a prompt search for $h\rightarrow ss \rightarrow 4b$ is also shown at small lifetimes.
} 
\label{fig:LLP_Jets}
\end{figure}

Searches for $h \rightarrow ss $ where $s$ is a new long-lived scalar in the mass range $30-40$~GeV decaying to SM fermions are summarized in Figure~\ref{fig:LLP_Jets}. For the SM+s scenario, decays to pairs of $b$-quarks dominate with $Br(s\rightarrow bb)\approx 85\%$ in this mass range, with additional significant decays to $c$-quarks $Br(s\rightarrow cc)\approx 5\%$ and tau leptons $Br(s\rightarrow \tau\tau)\approx 8\%$ (see Section~\ref{sec:benchmark_sm_plus_s}).

Several results only include an interpretation for a combination of final states while others provide interpretations in terms of exclusive final states. In order to compare the results, Figure~\ref{fig:LLP_Jets} shows results for $(\sigma_h/\sigma_h^{SM}) \times Br(h\rightarrow ss)$ using the approximate branching ratios for $s$ to decay to $b$, $c$, and $\tau$ final states quoted above. The ATLAS CalRatio, MS and ID+MS DV results~\cite{ATLAS:2018tup,ATLAS:2019jcm,ATLAS:2019qrr,ATLAS-CONF-2021-032} provide only the combined results and are reproduced without modifications. The curves corresponding to the ATLAS~\cite{ATLAS:2021jig} and CMS~\cite{CMS:2021uxj,CMS:2020iwv} displaced jets analyses use only the $bbbb$ final state with an assumed $Br(s\rightarrow bb) = 85\%$. The difference in per-jet efficiency for displaced jet reconstruction in decays to different quark flavors comes primarily from the difference in track multiplicity. Jets originating from $b$-quarks may form tertiary vertices, effectively reducing the number of tracks in the LLP decay vertex.  We expect per-jet efficiencies for decays to $c$-quarks to be more similar to those for light quark jets than $b$-jets; this effect is seen in 
the LHCb analysis~\cite{LHCb:2017xxn}. 
However, due to the limited information available, the final states with $s\rightarrow cc$ and $s\rightarrow \tau\tau$ decays are not included in the figure for ATLAS and CMS displaced jet results.

The LHCb displaced jet analysis~\cite{LHCb:2017xxn} includes results both in the $bbbb$ and $cccc$ final states, which are combined in Figure~\ref{fig:LLP_Jets} under the assumption that the only difference comes from the per-jet efficiency for displaced jet reconstruction.  
The CMS hadronic MS analysis~\cite{CMS:2021juv} provides results in the $bbbb$ and $\tau\tau\tau\tau$ final states and a similar combination is performed. Beyond the analyses using dedicated reconstruction methods for long-lived resonances, a reinterpretation of the ATLAS prompt search~\cite{ATLAS:2020ahi} shows sensitivity to signals with lifetimes up to the mm scale. This analysis is mainly sensitive to the $bbbb$ final state, since b-tagging algorithms are used to select events.

\subsubsection{SM+v}
\label{sec:exp_llp_sm_plus_v}

\begin{figure}[htb]
    \centering
        \includegraphics[width=\textwidth]{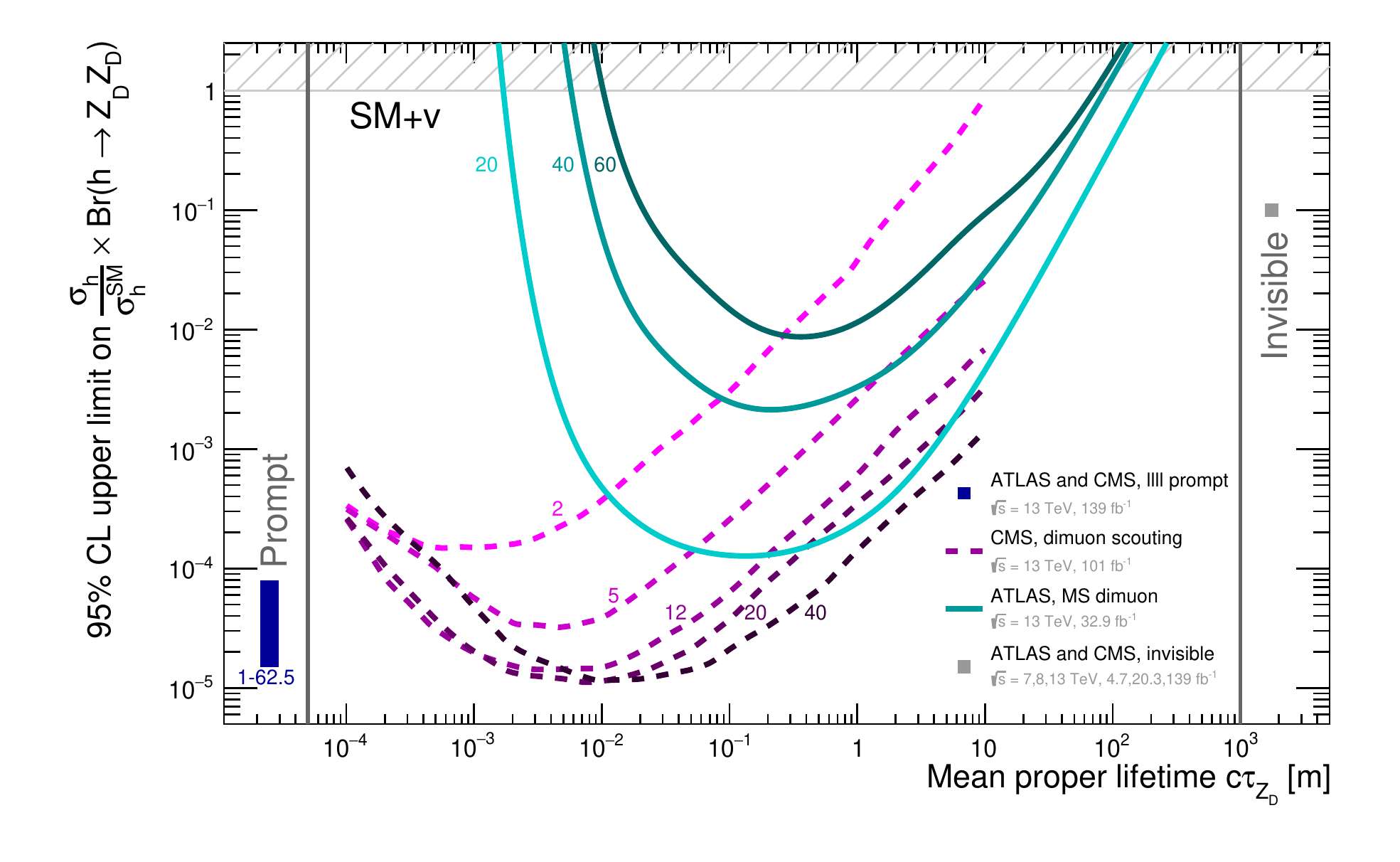}
\caption{Observed 95\% CL upper limits on $\sigma_h/\sigma_h^{SM} Br(h\rightarrow Z_{D}Z_{D})$ in the SM+v scenario where $Z_D$ is a new long-lived vector decaying to leptons. The analyses are summarized in Table~\ref{tab:llp_summary}. The branching fractions of $Z_D$ to SM particles follow the prescriptions in~\cite{Curtin:2014cca}, as described in Section~\ref{sec:benchmark_sm_plus_v}. Current limits for prompt searches (see Table~\ref{tab:prompt_summary}) and searches for invisible Higgs boson decays (see Table~\ref{tab:hinv_summary}) are also shown. The mass of the corresponding signal LLP is indicated next to each curve in GeV.}
\label{fig:LLP_Muons}
\end{figure}

Recent results targeting searches for exotic Higgs decays to LLPs decaying to leptons are summarized in Figure~\ref{fig:LLP_Muons}. These searches are interpreted in the context of the SM+v scenario, described in Section~\ref{sec:benchmark_sm_plus_v}, focusing on leptonic decays of a long-lived $Z_D$, $h\rightarrow Z_{D}Z_{D}$. 
The limits from prompt searches, described in Section~\ref{sec:exp_prompt}, and limits from invisible decays, shown in Table~\ref{tab:hinv_summary}, are also shown for comparison. A reinterpretation of these searches in terms of signals with small lifetimes is currently not available, but would be interesting to show the range of lifetimes covered by prompt searches and close the gap with dedicated LLP searches. 

Dedicated triggering techniques targeting displaced muons have been instrumental to search for LLP decays in the $2\mu$+X and $4\mu$ final states. A recent search by CMS~\cite{CMS:2021ogd} using dimuon scouting probes a wide range in proper lifetimes (0.1~mm to 10~m) for masses in the range 0.65-50~GeV and constrains branching ratio limits down to about $10^{-5}$. A complementary ATLAS~\cite{ATLAS:2018rjc} search for LLP $\mu\mu$+X decays, identified through displaced vertices reconstructed from tracks in the MS that do not have a corresponding ID track, extends the probed lifetime range beyond 100 meters, with branching ratio limits down to about $10^{-4}$ for lifetimes in the cm-m range and LLP masses of 20 GeV. These results provide lifetime exclusions for several mass hypotheses and also include interpretations as a function of other model parameters, such as the kinetic mixing and mixing parameter $\kappa$.      

Available searches offer sensitivity to a broad range of lifetimes for a range of masses. Extending the reach of the analyses to target lifetimes and masses beyond those currently reported in the interpretations is important to cover the full range of possible signals. Two-dimensional $m$-$c\tau$ plots may provide a clearer picture of any gaps in sensitivity. Several interpretations are coarse and do not cover the full region of sensitivity in mass and lifetime. In addition, limits for different exclusive final states may complete our understanding of the sensitivity of different searches and help guide future search efforts. 

The special techniques used in LLP searches are a challenge for the reinterpretation of results for alternative models without access to the full simulation of detector effects. Data preservation and in-depth documentation of the details of the models, including detailed acceptances and efficiencies, following the recommendations of~\cite{LHCReinterpretationForum:2020xtr}, is particularly important in this context. As an example, Ref. \cite{ATL-PHYS-PUB-2020-007} presents a reinterpretation of the ATLAS displaced calorimeter jet search~\cite{ATLAS:2019qrr} with the RECAST framework~\cite{Cranmer:2010hk} in the context of a dark SUSY model~\cite{Falkowski:2010cm,Falkowski:2010gv} that considers hadronic decays of a dark photon and provides complementary coverage to a dedicated search for displaced lepton-jets that targets such dark photon decays to collimated dimuons or light hadrons~\cite{ATLAS:2019tkk}. 

\section{Future probes}
\label{sec:future}

The HL-LHC is expected to produce a large sample of $\mathcal{O} (10^8)$ Higgs bosons, which will generically enable more than an order of magnitude improvement in the sensitivity to exotic Higgs boson decays. Projections assuming that current object reconstruction performance can be maintained indicate that sensitivities to exotic Higgs decays into hadronic final states can reach 
sub-percent branching fractions~\cite{CMS:2019rsy}. Meanwhile, combining the full ATLAS and CMS datasets from the HL-LHC  is forecast to constrain the total exotic branching ratio to $< 4\%$ at 95\% CL \cite{deBlas:2019rxi}.  The improved discovery potential from the HL-LHC is not only from the large total integrated luminosity (up to 3~ab$^{-1}$), but also the planned detector upgrades, which will provide refined trigger and reconstruction capabilities.

The detector upgrades include new inner tracking and timing detectors, as well as new electronics and an updated trigger and data acquisition system, which are expected to extend the capabilities in ATLAS and CMS to study exotic Higgs decays in challenging low-mass and/or displaced final states. Studies from CMS have shown that improvements in displaced jets using information from the ID and time-displaced calorimeter-based jets can significantly improve the sensitivity to LLPs with meter-scale lifetimes 
\cite{CERN-LHCC-2020-004}. Both ATLAS and CMS plan to reduce the dependence of muon triggers on primary vertex constraints, increasing the sensitivity to LLP decays with larger displacements~\cite{CERN-LHCC-2020-004,CERN-LHCC-2017-020}. The increased triggering efficiency for muons with larger displacements can also significantly improve the sensitivity to exotic Higgs decays to dark photons, such as the CMS HL-LHC prospect described in~\cite{CMS:2018lqx}. The higher granularity in the ATLAS triggers will also make it possible to identify collimated dimuon pairs arising in decays of low mass resonances~\cite{CERN-LHCC-2017-020}.

Beyond the main LHC program, there are several other experimental strategies being pursued or planned that provide complementary opportunities to study exotic Higgs boson decays. Several dedicated LLP detectors provide new capabilities for discovering LLPs produced in the LHC collisions (Section~\ref{sec:dedicated_llp_exp}). Future colliders offer substantial new opportunities for Higgs decays.  
Electron-positron and muon-muon colliders offer a clean environment to probe such signals (Sections~\ref{sec:ep_colliders}, \ref{sec:muon_colliders}, respectively), while high-energy proton colliders enable searches for extremely rare decays (Section~\ref{sec:had_colliders}), as we now discuss. 

\subsection{Dedicated LLP detectors}
\label{sec:dedicated_llp_exp}

Triggering challenges are one major limitation for search sensitivity to LLPs produced in the relatively low-energy  events following from Higgs decays.  Proposed dedicated LLP detectors, such as CODEX-b \cite{Gligorov:2017nwh,Aielli:2019ivi} and MATHUSLA \cite{Chou:2016lxi,MATHUSLA:2018bqv}, would be located far away from the collision point with intervening shielding material that would stop most of the charged and/or strongly-interacting SM particles from reaching the detectors.  These detectors would not suffer from trigger limitations and would thus readily have sensitivity to the relatively low mass final states of interest for LLPs produced in Higgs decays.  The long baselines and low backgrounds of these experiments also help them extending the range of  $c\tau$ that can be probed, although the small solid angle subtended by these detectors, compared to the main detectors, is the limiting factor in this extension.  For theories such as neutral naturalness models that predict a large ($\gtrsim 3$) number of relatively soft LLPs in a single event, the soft $p_T$ spectra for the LLPs tend to exacerbate triggering challenges at the main detectors, while the larger number of LLPs increases the acceptance at dedicated detectors.  In such scenarios the additional physics reach provided by dedicated detectors can be substantial \cite{Curtin:2018mvb}. 

The Higgs is sufficiently heavy that, to have optimal acceptance for its decay products, detectors should not be located too far forward.  MATHUSLA and CODEX-b have good acceptance for exotic Higgs decay products, while far-forward detectors such as the FASER experiment \cite{Feng:2017uoz,FASER:2018eoc} do not have competitive sensitivity.  
However, when new physics is light enough to be produced in e.g. meson decays, a wide range of new experimental signatures and searches become possible. In particular, exotic $B$ and $K$ decays can be powerful probes of BSM scalars coupled through the Higgs portal.  The relationship between such low-energy probes and exotic branching ratios of the Higgs boson is highly model-dependent, and we will not discuss the topic further here; see~\cite{Beacham:2019nyx} for a recent overview.

\subsection{Future colliders}

\subsubsection{Electron-positron colliders}
\label{sec:ep_colliders}

The $\mathcal{O}(10^6)$ Higgs bosons that would be produced at planned electron-positron colliders will not come close to the statistics of the HL-LHC Higgs sample, but thanks to the low-background environment, will 
offer the most sensitive discovery opportunities for decays into hadronic final states, as well as other final states that cannot be easily reconstructed and/or separated from the large LHC backgrounds.  This Higgs sample size is reflective of the running scenarios summarized in \cite{deBlas:2019rxi} for three proposed machines: 5 (5.6) ab$^{-1}$ of unpolarized collisions at 240 GeV, as envisioned for FCC-ee \cite{FCC:2018byv} (CEPC \cite{CEPCStudyGroup:2018ghi}), and 2 ab$^{-1}$ of polarized collisions at 250 GeV, which is the first stage of operation planned for the ILC \cite{Bambade:2019fyw,LCCPhysicsWorkingGroup:2019fvj}.

A survey of electron-positron collider sensitivities for several (partially-)visible, prompt final states was performed in \cite{Liu:2016zki}, which found sensitivities to exotic branching ratios at the level of a few$\times 10^{-4}$ over a wide range of signatures.  These studies were carried out for the particularly clean process where $e^+e^-\to Zh$ is followed by $Z\to\ell\ell$. Further incorporating analyses in the higher statistics but also higher-background $Z\to jj$ or $Z\to$ invisible final states would contribute additional sensitivity. 
Meanwhile direct searches for invisible decays are projected to reach $Br (h\to \mathrm{invisible}) < 0.3\%$ \cite{deBlas:2019rxi}. 

The SM+s model is a prime example of a model that predict preferential decays to final states that have challenging backgrounds at the LHC.
HL-LHC sensitivity to SM+s above the $b$ threshhold reaches up to $Br(h\to ss) < 2.8\times 10^{-2}$ using CMS HL-LHC projections for the $bb\tau\tau$ final state \cite{CMS:2019rsy}.  This projected sensitivity is  exceeded by nearly two orders of magnitude  at  electron-positron colliders: the estimates of Ref.~\cite{Liu:2016zki} for the $bbbb$ final state give $Br(h\to ss) \lesssim 5\times 10^{-4}$ across the range of masses $20\gev < m_s <60$ GeV.

Beyond dedicated exclusive searches, the known initial state at lepton colliders also affords the possibility of unambiguously measuring the total width of the Higgs boson, thereby allowing for fully model-independent upper bounds on the total exotic branching fraction.
Ref.~\cite{deBlas:2019rxi} projects the resulting 95\% CL constraints on the Higgs exotic branching ratio at the one percent level. 

While LLP searches at the LHC often enjoy excellent background rejection at the analysis level, trigger challenges can limit LHC efficiency for LLPs produced via the relatively low-mass Higgs boson.  The lower background environments of lepton colliders, together with detectors constructed with BSM LLP signatures in mind, can potentially make LLP searches at Higgs factories competitive with those at the LHC at the shorter LLP lifetimes where LHC backgrounds are higher \cite{Alipour-Fard:2018lsf}, or in other scenarios where the event presents a particular trigger or background rejection challenge at the LHC.


\subsubsection{Future hadron colliders}
\label{sec:had_colliders}

Future proton-proton colliders running at a higher center of mass energy will produce a much larger sample of Higgs bosons. For example, $\mathcal O(10^9)$ Higgs bosons will be produced at the high-energy LHC ($\sqrt s=27$ TeV) with $15\,\mathrm{ab}^{-1}$ data and at a 100 TeV proton collider with $3\,\mathrm{ab}^{-1}$ data. 
This very large sample of Higgs bosons can give access to very rare exotic decay modes. The biggest improvements on the reach will be obtained for very clean decay modes, which particularly benefit from the increase in statistics. An example is the search for $h\to Z_DZ_D\to 4\ell$.
The bound on the branching ratio for $h\to Z_DZ_D$ could be pushed to $\sim 10^{-7}$, more than an order of magnitude more stringent bound than what is achievable at the HL-LHC \cite{Curtin:2014cca}. Another example is the Higgs decay into two long-lived dark glueballs, as arising in neutral naturalness theories \cite{Curtin:2015fna}. Meanwhile direct searches for the invisible branching fraction will reach $Br(h\to\inv) < 2.5\times 10^{-4}$, well below the SM prediction $Br(h\to  ZZ^*\to 4\nu) = 1.1\times 10^{-3}$ \cite{FCC:2018vvp}. Global fits of Higgs rates will constrain the branching fraction into untagged states below $1\%$~\cite{deBlas:2019rxi}.

However, direct searches for higher-background exotic Higgs decay modes can also see major advances in sensitivity at future high energy proton colliders. Cross-sections for sub-leading Higgs production modes can be much larger at higher-energy proton colliders than the corresponding cross-sections at the LHC. For example, the $t\bar t h$ production cross section at a 100 TeV proton collider is $\sim 60$ times larger than at the LHC. These subleading production modes will be essential to improve the reach for background-limited exotic Higgs decays, such as e.g. $h\to ss\to bbbb$. 

\subsubsection{Muon colliders}
\label{sec:muon_colliders}

Muon colliders, while more challenging to construct and instrument, would be able to realize high center-of-mass energy collisions between  elementary particles. 
Although potential operational scenarios are still under active discussion, the inclusive Higgs production at an $\mathcal O(10)$ TeV muon collider with $\mathcal O(100 \,\mathrm{ab}^{-1})$ of data would yield $\mathcal{O}(10^8)$ Higgs bosons, two orders of magnitude larger than the production at proposed $e^+e^-$ Higgs factories. In addition, given the low-background environment at a muon collider, such machines can provide an excellent opportunity to investigate exotic Higgs decays beyond the reach of the HL-LHC, offering sensitivity to ultra-rare exotic decay modes with branching fractions down to $\mathcal O(10^{-7})$ \cite{AlAli:2021let,Franceschini:2021aqd}.

\section{Summary and conclusions}
\label{sec:conclusions}

Exotic decays of the Higgs boson are an excellent place to search for the footprints of new physics that couples only feebly to the SM. In the years since the discovery of the Higgs boson, many theoretical studies of and experimental searches for exotic Higgs boson decays have been performed.
This review summarizes recent developments in theoretical motivations for exotic Higgs decays as well as the status of experimental searches for signatures involving both prompt and displaced final states.  We present interpretations of many searches in terms of three benchmark scenarios that represent some of the best-motivated minimal models for exotic Higgs decays. 

The searches reviewed here already constitute a substantial accomplishment.  Experimental searches for dark vector bosons have attained sensitivity to unprecedentedly small Higgs branching fractions, and even searches to more difficult final states are reaching notable landmarks. Direct LHC searches for exotic decay modes are now the leading probes of many models.  Even for the challenging hadrophilic SM+s model, direct searches in a variety of both prompt and displaced final states now reach beyond constraints on the total exotic branching ratio imposed by global fits to Higgs properties, while searches for the decay $h\to Z Z_D$ are now beginning to
surpass precision electroweak measurements as a probe of the minimal dark photon model in parts of parameter space.
On the theoretical side, a substantial body of recent work has led to qualitatively new understanding of how exotic Higgs decays can be related to unanswered questions in particle physics, most notably the hierarchy problem and the origin of dark matter.

To build on this effort and realize the full discovery potential offered by exotic Higgs decays,  work is still needed to understand  gaps in current experimental coverage, to guide future searches at the LHC and beyond, and to use our developing understanding of the Higgs boson to shed new light on physics beyond the SM. 
A major outstanding frontier for both theory and experiment is Higgs decays involving more than one BSM species, as well as the extension of current searches into more challenging regimes.

BSM particles are motivated across the entire  mass range accessible in Higgs decays, as well as with any experimentally testable value of $c\tau$.
Many current searches can be extended to cover a broader range of masses.  In the lower mass regime, the use of reconstruction and identification techniques targeting non-isolated or merged decay products can be used, also in combination with resolved objects to extend the reach. Existing searches for Higgs decays may also offer sensitivity at the lowest masses where the decay products are completely overlapping.  Dedicated strategies may narrow gaps, including the $J/\psi$ and $\Upsilon$ regions.

The further development of LLP searches will be critical for extending coverage across the full range of testable lifetimes.  Reinterpretations of prompt searches are important to identify gaps in search coverage for the smallest $c\tau$. 
Extending sensitivity at long lifetimes may be possible by focusing on the outermost layers of current detectors and exploring dedicated LLP detectors.  Searches for LLPs can be harder to reinterpret than prompt searches; thus it is vital for searches for decays to LLPs to report their results in as transparent and reusable a form as possible.  This point is especially pressing as in many well-motivated theories the LLPs are composite states, whose production is challenging to model in detail.  

Many final states of interest are currently not covered.
 Examples of uncovered final states include fully hadronic decays to jets (e.g., $gggg$,  $ggbb$, etc.), or displaced decays to photons or taus.  Thus direct searches are not currently sensitive to Higgs decays involving hadrophilic ALPs that decay dominantly to gluons, nor do they target the interesting case of the displaced photonically-decaying ALP.
Meanwhile, few searches for semi-visible final states have been performed to date, or for decays including a $Z$ boson and a new boson in the decay.  More complex scenarios involving decays into two different BSM particles, including the case where cascade decays can occur among the BSM species, have not been extensively explored.  It does not take a particularly elaborate model to realize high multiplicity final states; the simple Higgsed dark photon model can already realize this possibility.
In many cases of interest high-multiplicity final states readily arise from decays into confining hidden sectors.  Given the experimental challenges posed by soft, high-multiplicity final states together with the enlarged parameter space of multi-state models, theoretical work on motivated reference models featuring multiple BSM species can be particularly valuable to help guide this program.  However some signatures such as semi-invisible decay modes (very common in decays to confining hidden sectors) and the potential presence of resonances at different masses are straightforward and can be readily implemented.

Exotic Higgs decays continue to serve as strong motivation for future work. Sensitivity studies, including considerations about dedicated triggers and reconstruction strategies for these signals, can provide valuable input to guide future search strategies. Theoretical and phenomenological work continues to be crucial to understand the implications of these analyses for physics beyond the SM and to set priorities for future efforts. The LHC and its HL-LHC upgrade, as well as future colliders and experiments, will offer many exciting opportunities to explore the question of whether the Higgs has new interactions and help answer some of the outstanding open questions in particle physics.

\section*{Acknowledgements}
The work of MC is supported by the Spanish Ministry of Science and Innovation ``Ram\'on y Cajal'' fellowship RYC-2017-22161. The research of SG is supported in part by NSF
CAREER grant PHY-1915852 and grant  PHY-1748958. 
SG thanks the Aspen Center for Physics, supported by NSF grant PHY-1607611, for hospitality when part of this work was done. 
The work of VM is supported in part by the NSF CAREER grant PHY-1830832.  The work of JS is supported in part by DOE Career grant DE-SC0017840. We thank J. Stegemann, N. Wardle, A. Escalante, C. Botta, R. Coelho Lopes de Sa and S. Willocq for useful discussions.

\bibliography{exohreview}
\bibliographystyle{JHEP}

\end{document}